\documentclass[preprint2]{aastex6}

\usepackage{color}
\usepackage{amsmath}
\usepackage{natbib}
\usepackage{graphics,graphicx}
\usepackage{wrapfig}
\usepackage{rotating}
\usepackage{gensymb}
\usepackage{fancyref}
\usepackage{moresize}
\usepackage{tabularx}

\newcommand{\es}{ergs s$^{-1}$}

\shortauthors{Johnson et al. (2017)}

\begin{document}


\title{Using Strong Gravitational Lensing to \\Identify Fossil Group Progenitors}


\author{Lucas E. Johnson\altaffilmark{1}, Jimmy A. Irwin\altaffilmark{1,2}, Raymond E. White III\altaffilmark{1}, Ka-Wah Wong\altaffilmark{3,4}, W. Peter Maksym\altaffilmark{5}, Renato A.  Dupke\altaffilmark{4,6,7}, Eric D. Miller\altaffilmark{8}, \& Eleazar R. Carrasco\altaffilmark{9}}


\altaffiltext{1}{Department of Physics \& Astronomy, University of Alabama, Box 870324, Tuscaloosa, AL 35487, USA; e-mail: lejohnson4@crimson.ua.edu}
\altaffiltext{2}{Department of Physics \& Astronomy, Seoul National University, Seoul 08826, Korea}
\altaffiltext{3}{Department of Physics \& Astronomy, Minnesota State University, Mankato, MN 56001, USA}
\altaffiltext{4}{Eureka Scientific Inc., 2452 Delmer St. Suite 100, Oakland, CA 94602, USA}
\altaffiltext{5}{The Harvard-Smithsonian Center for Astrophysics, 60 Garden St., MS-67 Cambridge, MA 02138}
\altaffiltext{6}{Department of Physics \& Astronomy, University of Michigan, 450 Church St., Ann Arbor, MI 48109, USA}
\altaffiltext{7}{Observat\'{o}rio Nacional, Rua Gal. Jos\'{e} Cristino 77, S\~{a}o Crist\'{o}v\~{a}o, CEP20921-400 Rio de Janeiro RJ, Brazil}
\altaffiltext{8}{Kavli Institute for Astrophysics \& Space Research, Massachusetts Institute of Technology, 77 Massachusetts Ave., Cambridge, MA 02139, USA}
\altaffiltext{9}{Gemini Observatory/AURA, Southern Operations Center, AURA, Casilla 603, La Serena, Chile}


\begin{abstract}
{Fossil galaxy systems are classically thought to be the end result of galaxy group/cluster evolution, as galaxies experiencing dynamical friction sink to the center of the group potential and merge into a single, giant elliptical that dominates the rest of the members in both mass and luminosity.  Most fossil systems discovered lie within $z<0.2$, which leads to the question: what were these systems' progenitors?  Such progenitors are expected to have imminent or ongoing major merging near the brightest group galaxy (BGG) that, when concluded, will meet the fossil criteria within the look back time.  Since strong gravitational lensing preferentially selects groups merging along the line of sight, or systems with a high mass concentration like fossil systems, we searched the CASSOWARY survey of strong lensing events with the goal of determining if lensing systems have any predisposition to being fossil systems or progenitors.  
We find that $\sim$13\% of lensing groups are identified as traditional fossils while only $\sim$3\% of non-lensing control groups are.  We also find that $\sim$23\% of lensing systems are traditional fossil progenitors compared to $\sim$17\% for the control sample.  Our findings show that strong lensing systems are more likely to be fossil/pre-fossil systems than comparable non-lensing systems.  Cumulative galaxy luminosity functions of the lensing and non-lensing groups also indicate a possible, fundamental difference between strong lensing and non-lensing systems' galaxy populations with lensing systems housing a greater number of bright galaxies even in the outskirts of groups.}
\end{abstract}





\section{Introduction}
Fossil systems are classically thought to be representative of old, undisturbed galaxy systems where almost all $L^*$ members have been cannibalized by the dominant central elliptical, as dynamical friction draws the massive $L^*$ member galaxies to the center over long time scales.  As the central elliptical cannibalizes more galaxies, the magnitude gap between the central elliptical and the next most massive member widens and mass becomes more concentrated at the center until a `fosssil system' is created (Jones et al.\ 2003).  One study of the mass concentration of fossil groups using N-body simulations show some support for this assumption toward fossil systems formation  (Khosroshahi et al.\ 2007).  Due to the apparent evolved nature of fossil systems and high concentration parameters (Wechsler et al. 2002), it is possible that studying fossil systems can help us understand of properties of the early universe as well as brightest cluster/group galaxy formation.  However, exactly how common fossil systems are in the universe is not well constrained.

Jones et al.\ (2003) originally defined a fossil system to be a galaxy group or cluster with a massive brightest group or cluster galaxy (BGG or BCG) that dominates (by more than two magnitudes in the $r$-band) the rest of the galaxies within $0.5R_{200}$, defined as half of the virial radius of the system, and shows a bolometric X-ray luminosity $L_{X, bol}\geq10^{42}h^{-1}_{50}\;$\es.  This definition has done well in identifying  massive fossil systems but has the potential to miss poorer fossil groups along with being less robust for these poor fossil groups (Dariush et al.\ 2010), since the infall of a lone luminous galaxy would destroy the system's fossil status.  To address this, Dariush et al.\ (2010) proposed altering the classic Jones optical criteria: instead of using ($\Delta m_{12}\geq2.0$), where $\Delta m_{12}$ is the $r$-band magnitude gap between the first and second rank galaxies, Dariush et al.\ (2010) requires ($\Delta m_{14}>2.5$), where $\Delta m_{14}$ is the $r$-band magnitude gap between the first and fourth rank galaxies.  This change would allow for the infall of one or two luminous galaxies without destroying the fossil status of the poorer group.  Jones et al. (2003) found that fossil systems should comprise between 8\% and 20\% of all galaxy groups.  However, this study was only done for nearby groups and clusters of $z<0.25$.  A more recent study by Gozaliasl et al.\ (2014)  shows that the fossil group fraction for massive galaxy groups ($M_{200}\sim10^{13.5}\;$M$_{\odot}$), where $M_{200}$ is the mass within a sphere of density equal to 200 times the critical density of the universe, is  $22\pm6\%$ for $z\leq0.6$ and $13\pm7\%$ for $0.6<z<1.2$ if a $\Delta m_{12}\geq1.7$ is required. 


Since most fossil systems found to date lie within $z<0.2$, fossil galaxy groups could be old, undisturbed systems, as the infall of any bright galaxies has the potential to destroy their fossil statuses.  The Millennium Simulation supports this idea, as it shows fossils being formed near $z=0.9$ and subsequently being destroyed due to the infall of a bright $L^*$ galaxy which breaks the $r$-band magnitude gap requirement before $z=0$ (von Benda-Beckmann et al. 2008).  However, as some fossils are destroyed in the simulation, others are created as bright members are consumed by the BGG (Ponman et al.\ 1994) thus establishing the required $r$-band magnitude gap.  This result from the simulation suggests that fossil systems could be more of a `fossil phase' that all groups have a likelihood of passing through as opposed to a unique class of object all their own.  There also exists controversial evidence that fossils have a higher than expected mass concentration parameter (Sun et al.\ 2004; Khosroshahi et al.\ 2004, 2006), although some observations of nearby fossils dispute this (Sun et al.\ 2009).  These pieces of evidence point to fossil systems possibly having different initial conditions than most groups.  To further complicate the matter, there are many supposedly old, evolved nearby fossil systems that do not possess cool cores as would be expected from relaxed systems (Sun et al. 2004; Khosroshahi et al.\ 2004, 2006).  These fossil systems needed an energetic event of some kind (such as an AGN turning on or a group merger) in their recent past to heat up their intragalactic mediums (IGM) or destroy any cool cores which is at odds with these fossils being relaxed.

Despite the certain existence of fossil progenitors, little observational work has gone into locating any.  A progenitor to today's fossil systems would be a system at a higher redshift with ongoing or imminent major merging, in mid-assembly of the eventual fossil system's massive BGG\footnote[1]{It is important to note that it is logically possible for a BGG/BCG to be extremely efficient at turning gas into stars, essentially preventing any other large galaxies from forming in the group/cluster.  Fossil systems formed via this channel would simply be ``born that way". }.  The amount of merging concluded by $z=0$ would be sufficent to push the final BGG $r$-band magnitude two magnitudes brighter than any other remaining galaxy member within $0.5R_{200}$.  The progenitor could also be more centrally mass concentrated than other non-fossil groups at similar redshifts if fossil systems are a unique set of galaxy groups as some suggest.  We expect fossil progenitors to be distinctly different from compact groups, since compact groups do not show velocity dispersions indicative of a deep cluster-like potential well, and the formation of fossil systems is not explained by the merger of galaxies in compact groups (La Barbera et al.\ 2010).

An overarching question in the study of fossil systems is what these groups looked like in the early universe.  Are they all old, evolved systems, the inevitable end for all clusters, or are fossils simply a phase that all groups have a probability of transitioning through?  While the former explanation is possible, the likelihood that the entire fossil population is comprised of isolated, undisturbed groups seems low based on the frequency of mergers in the early universe as seen in simulations.  Studies of the Millennium Simulation also cast doubt on this being the sole explanation.  Many fossil systems in the simulation form between $0.3<z<0.6$, and the fossil system BGGs were always larger than their non-fossil counterparts (Kanagusuku et al.\ 2016), suggesting that fossil systems could begin with different initial conditions than most galaxy systems.  By studying the Cheshire Cat fossil progenitor system, it has been demonstrated that if two groups merge, the final product has the potential to be a fossil group once the BGGs of the respective groups merge (Irwin et al. 2015).  It was estimated that the first and second rank galaxies in the Cheshire Cat gravitational lens will merge in 0.9 Gyr, and once this merger concludes, the group will become a fossil system.  Further, optical and X-ray observations revealed that the system is comprised of two separate galaxy groups undergoing a line of sight merger, opening another avenue for fossil system formation: (fossil) group mergers.  This second possible formation mechanism offers an explanation for observed non-cool core fossil systems (as X-ray cooling time scales are typically longer than galaxy merger time scales.)

It is known that fossil systems house massive BGGs at their centers (Jones et al.\ 2003), implying a higher than average mass concentration when compared to normal groups of similar richness.  This enhancement could make fossil-like systems more efficient stong gravitational lenses.  Since gravitational lensing also preferentially selects merging systems along the line of sight, it follows that targeting systems with strong gravitational arcs nearby could be a more efficient way of locating fossil systems and their progenitors.  Kanagusuku et al.\ (2015) found that in the Millennium Simulation, most of today's fossils became fossils between $0.3<z<0.6$ which happens to be the optimal distance to observe groups act as strong lenses (Trentham 1995).  This motivates us to select our sample from the CAmbridge Sloan Survey of Wide ARcs in the Sky (CASSOWARY)  catalog (Belokurov et al.\ 2009, Stark et al.\ 2013) which searches for strong gravitational arcs in the Sloan Digital Sky Survey (SDSS) and has an average lens redshift of $z\sim0.4$.  Our goals in this study are to identify more examples of fossil progenitors (such as the Cheshire Cat) in the CASSOWARY catalog using SDSS photometry, form a catalog of these progenitors, and contrast their properties against fossil and non-fossil systems.  A control set of near-identical, non-lensing galaxy groups will also be analyzed to see if the presence of a strong gravitational arc near a group biases it toward being a fossil system.  We present results from our analysis of all 58 CASSOWARY members along with average cumulative luminosity functions for each category (fossil, progenitor, and normal systems).

In Section 2, we discuss the selection criteria for our sample from the SDSS archive, group scaling relations involved in determining each group's physical parameters, how fossil status is determined, and how average luminosity functions were generated.  Section 3 presents our catalog of lensing fossil systems and potential fossil progenitor systems and compares the results against a control sample of near identical, non-lensing galaxy groups.  In Section 4, we propose a possible fossil system formation timeline using data from SDSS, incorporating fossil progenitors at varying stages of BCG/BGG formation.  Section 5 summarizes our findings.  We adopt the standard  $\Lambda$CDM cosmology with $H_0=70\;$km s$^{-1}$ Mpc$^{-1}$ and $\Omega_{M}=0.286$ throughout this work.

\section{Sloan Data Analysis}
\subsection{\textbf{Selection Criteria}}



The CASSOWARY catalog (Belokurov et al.\ 2009, Stark et al.\ 2013) identified strong gravitational arcs in the SDSS DR7 archive by searching for blue companions or arcs separated from a luminous red galaxy by $\sim$3".  To date, 58 lensing systems have been identified with many having been confirmed via spectroscopic observations of both the lensing and lensed galaxies with typical lensing galaxies lying between $0.2<z<0.7$ and lensed galaxies beyond $z\sim1.5$.  While it was not required that the lens be a galaxy group, it was found that most were.  Moreover, many of the CASSOWARY groups were not previously identified in group catalogs, as they generally have few members and their higher redshifts create difficulties for automated selection techniques.

\begin{figure}[h]
\vspace{-1.25truein}
\hspace{-0.2truein}
{\includegraphics[scale=0.4625]{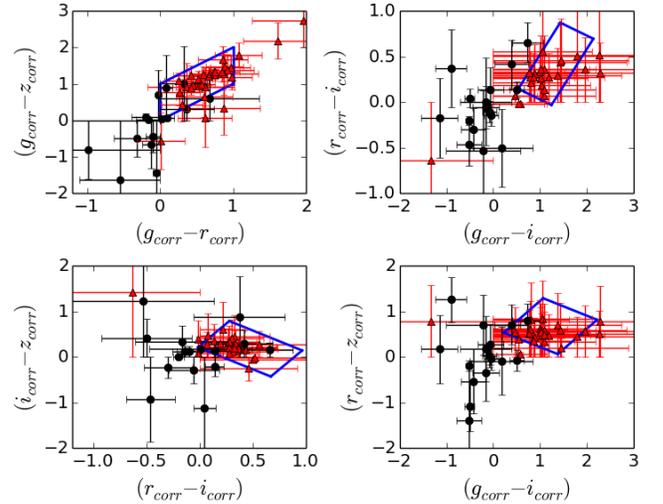}}
\vspace{-1.25truein}
\caption{\footnotesize
{The four color-color group inclusion parellograms created for the Cheshire Cat (CSWA 2) for red-ridge ellipticals;  K-corrections and stellar evolutionary corrections have been included. Red triangles show confirmed member galaxies and black circles show interlopers.  Error bars were calculated quadratically from the reported SDSS DR12 magnitude errors. Confirmed member galaxies outside the regions are very faint and were only confirmed via spectroscopic redshifts.}}
\label{fig:colors}
\end{figure}

Our analysis primarily used SDSS DR12 photometric data, which consists of over 100 million cataloged sources observed in the $ugriz$ bands that span over a quarter of the sky.  Unfortunately, due to the vastness of the data set, only a few percent of the brighter galaxies have available spectra, limiting our ability to know precise distances and therefore accurate group membership.  We therefore use photometric redshifts (photoZ) estimates provided in the SDSS archive which are found using the observed colors of the sources and correlate them to a database of spectroscopic redshifts (specZ) of galaxies of similar colors.  This method allows for an estimate of the source's distance, bearing in mind that improvements could be made once a spectrum is taken.

It is important to note where the uncertanities in SDSS's photoZ measures come from and how they can be minimized.  The error in the photoZ estimate directly correlates with the errors in the source's colors, meaning the fainter sources have less accurate corresponding photoZ.  Normally, at the distances of our targets, solely relying on photoZ to determine group membership is not optimal, so we developed a technique to construct reliable photoZ cuts for the other groups without spectroscopic information.  {\it Gemini} GMOS optical spectroscopic redshifts for 48 galaxies for the Cheshire Cat (also a CASSOWARY object known as CSWA 2) were available (Irwin et al. 2015) along with supplemental redshifts found in the literature; these allowed us to confidently determine group membership.  We therefore chose to use this group to construct our photometric inclusion criteria for the other CASSOWARY catalog systems.  By querying SDSS for the photoZ of each specZ confirmed member of the Cheshire Cat fossil progenitor and generating a histogram in photoZ space, we were able to quantify the spread in the photoZ distribution of known member galaxies.  We fit the distribution to a gaussian and found its standard deviation.  Using the photoZ errors provided by SDSS DR12, we determined that a window of $2\sigma$ centered on the peak of the distribution was sufficient to include all confirmed member galaxies.  If a galaxy's photoZ error bars fall within the $2\sigma$ inclusion window it was counted as a potential member and moved on to the next set of member cuts.

In order to count red-sequence (or red-ridge) elliptical galaxies for use in group scaling relations, we incorporate a series of color cuts to exclude blue, late-type galaxies as well as any red-ridge ellipticals at the wrong redshift.  The average redshift of the CASSOWARY groups, combined with most CASSOWARY member galaxies being red-ridge ellipticals, drove us to omit the $u$-band; the SDSS $u$-band errors were large for even the brightest BGG in our sample suggesting that including the $u$-band would add little information to our analysis.  Thus, every color combination using the $griz$ bands was inspected for the confirmed Cheshire Cat member galaxies.  The tighest groupings would limit the number of interlopers being accidentally counted among the red-ridge elliptical members (Figure~\ref{fig:colors}).  Four parallelograms in color-color space were chosen and are displayed in Table~\ref{tab:cuts}.  Galaxies outside any parallelogram are excluded.  These two criteria were used to determine group membership status for the rest of the CASSOWARY catalog.  The lack of spectra for each group means that when using these cuts the results will not be perfectly pure, however statistically it is complete.

To account for the relatively high redshift of the CASSOWARY groups,  K-corrections for groups of $z<0.5$ were found using the ``K-corrections Calculator" (Chilingarian et al.\ 2010)\footnote[2]{http://kcor.sai.msu.ru/ }.  For groups with $z>0.5$, K-corrections were estimated using $D_{n}4000$ measures (Westra et al. 2010) assuming a typical elliptical galaxy $D_n4000\sim1.65$.  Stellar evolutionary corrections for the $griz$ bands were taken from Roche et al.\ (2009).  Finally, we imposed a corrected $i$-band magnitude limit of $m_i\;<\;20.84$  to find $N_{200}$, defined as the number of red-ridge ellipticals with $L_{gal}>0.4L^*$ inside the group's virial radius, taken to be $R_{200}$ (Wiesner et al.\ 2012).

In order to find each group's $R_{200}$ and other physical properties, we adopted an iterative process using group scaling relations involving $N_{200}$ in Lopes et al. (2009), 
\begin{equation}
\textnormal{ln}(R_{200})=0.05+0.39 \textnormal{ln}(\frac{N_{200}}{25})h_{70}^{-1} \;\textnormal{Mpc}
\end{equation}
\begin{equation}
	\textnormal{ln}(M_{200})=0.21+0.83 \textnormal{ln}(\frac{N_{200}}{25})h_{70}^{-1} \; \textnormal{M}_{\odot}
\end{equation}
We first adopted a characteristic $R_{200}$ of $1\;$Mpc as a starting point. 
This size was used in an SDSS query to extract all galaxies within the circular angular region.  These galaxies were analyzed using our color/photoZ selection criteria which gave us a preliminary value for that system's $N_{200}$.  To deal with any remaining interlopers present within the extraction region, which could slightly inflate $N_{200}$ and consequently $R_{200}$, we took eight regions of angular size equal to the group's current $R_{200}$ immediately adjacent to the target group and applied the same selection process to the galaxies within these regions.  We averaged these eight results together to obtain an expected $N_{200}^{interloper}$ for each group based on the group's location in the sky.  This value was subtracted from the group's $N_{200}$ to arrive at a more accurate richness.  The new $N_{200}$ was used to calculate a new $R_{200}$, and the process was repeated until a single value for $N_{200}$ and $R_{200}$ was converged upon.  By using this method, we were able to agree with other SDSS galaxy cluster richness catalogs (Wen et al. 2012) on $N_{200}$ to within 10\% for the 17 groups with existing $N_{200}$ estimates.  Errors in galaxy colors were dealt with by adding in quadrature, and if a galaxy's color error bars pushed it into the inclusion regions, it was counted as a potential group member; this only significantly change results in the most distant groups.
\smallskip


\begin{table*}
{\scriptsize
\begin{tabularx}{\textwidth}{l c c c c}
\hline
\hline
 & Cut 1 & Cut 2 & Cut 3 & Cut 4 \\ 
\hline
R1&$(g-z)>-3.63(g-r)+3.0$&$(g-z)<1.05(g-r)+1.11$&$(g-z)<-3.63(g-r)+7.45$&$(g-z)>1.05(g-r)+0.35$ \\ 
R2&$(r-i)>-0.25(g-i)+0.28$&$(r-i)<0.81(g-i)-0.29$&$(r-i)<-0.25(g-i)+1.23$&$(r-i)>0.81(g-i)-1.04$ \\ 
R3&$(i-z)>-0.94(r-i)+0.20$&$(i-z)<1.95(r-i)+0.25$&$(i-z)<-0.94(r-i)+1.05$&$(i-z)>1.95(r-i)-1.73$ \\ 
R4&$(r-z)>-0.41(g-i)+0.63$&$(r-z)<0.90(g-i)+0.33$&$(r-z)<-0.41(g-i)+1.73$&$(r-z)>0.90(g-i)-1.18$ \\
\end{tabularx}
}
\caption{\footnotesize{Equations for lines used to build parallelograms in color space to exclude all galaxies except red-ridge ellipticals useful for group scaling relations with each inclusion region denoted by R1, R2, R3, and R4.  K-corrections and stellar evolutionary corrections are included to bring galaxies to a $z=0$ frame.}}
\label{tab:cuts}
\end{table*}

\subsection{\textbf{Determining Fossil Status}}

With color cuts, photometric redshift cuts, galaxy interloper averages, and reliable $R_{200}$ estimates in hand, we constructed galaxy membership catalogs for all 58 CASSOWARY members.  At this point, we checked the fossil status of each group using both the Jones et al. (2003) criteria of $\Delta m_{12}\geq2.0$ and the Dariush et al. (2010) criteria of $\Delta m_{14}\geq2.5$; systems which satisfied the optical criteria as-is were labeled as fossil systems.  Groups which were not classified fossil systems moved on to the next stage of analysis to determine whether they are fossil progenitors.  

Taking the brightest galaxy as the center of the system, we calculated the projected separation of all members from the BGG along with their masses (assuming a mass-to-light ratio of six in the $r$-band).  We took this information and calculated an expected time scale for the galaxy to merge with the BGG for each member using $T_{merge} \approx1.6r_{25}M_{*}^{-0.3}\;$Gyr, where $r_{25}$ is the maximum projected separation of the galaxies in units of $35.7h_{0.7}^{-1}\;$kpc and $M_{*}$ is the sum of the galaxies' masses in units of $4.3\times10^{11}h_{0.7}^{-1}M_{\odot}$ (Kitzbichler \& White 2008).   We adopt this time scale since it likely overpredicts the merger time by a factor of two relative to other methods (Kitzbichler \& White 2008) ensuring the merger is most likely completed in the specified time scale.   Based on the group's look back time, we determined which member galaxies had sufficient time to be cannibalized by the BGG and their luminosities added to the BGG.  Using this `new' BGG luminosity, the fossil status of each group was again checked; if a group became a fossil via this process it was labeled as a fossil progenitor.  Finally, all groups that still did not meet either fossil criteria were labeled for this work as normal groups, as no amount of possible merging could build a large enough BGG for these groups to transition into fossil groups by $z=0$ (Table 2).
\smallskip

\subsection{\textbf{Luminosity Functions}}

Once all 58 CASSOWARY catalog members were sorted based on their fossil status, we converted galaxy apparent $r$-band magnitudes ($m_r$) into absolute magnitudes ($M_r$) using
\begin{equation}
	M_r=m_r-25-5{\textnormal{log}}(\frac{D_L}{1\;{\textnormal{Mpc}}})-K_r-0.85z
\end{equation}
where $D_L$ is the luminosity distance to the group, $K_r$ is the $r$-band K-correction, and the last term is the stellar evolutionary correction involving the group's redshift (Roche et al. 2009).  We then generated three average luminosity functions (one for fossils, one for progenitors, and one for normal groups) using all the member galaxies for each category.  
Due to the low galaxy count in the brightest bins of the average luminosity functions, errors in galaxy count were handled using Poisson statistics with $\sigma\approx1+(n+0.75)^{\frac{1}{2}}$ where $n$ is the number of member galaxies within the luminosity bin (Gehrels 1986).
\smallskip

\section{Discussion}

\smallskip
\subsection{\textbf{CASSOWARY Strong Lensing Sample}}

Of the 58 CASSOWARY members, it was found that six are most likely large, lone ellipticals (possessing an $N_{200}<5$) that happen to act as strong gravitational lenses and were not included in any luminosity functions or fossil/progenitor fractions.  Of the remaining 52 strong lensing systems, we found that $13.5\pm2.8$\%\footnote[3]{Errors are reported at 1$\sigma$ confidence.}  are Jones fossils ($\Delta m_{12}\geq2.0$) and $17.3\pm2.6$\% are Dariush fossils ($\Delta m_{14}\geq2.5$), consistent with the expected 8\% to 20\% fossil system rate for randomly selected groups within $z<0.2$ (Jones et al. 2003).  We found that $23.1\pm2.5$\% of the CASSOWARY systems are Jones fossil progenitors and $28.9\pm2.5$\% are Dariush fossil progenitors.  This higher rate of fossil progenitors in the CASSOWARY sample is not surprising considering that Kanagusuku et al. (2016) found that in the Millennium Simulation, systems which were fossils at $z=0$ finished forming their BGG (creating the required $\Delta m_{12}$/$\Delta m_{14}$  $r$-band magnitude gap) between $0.3<z<0.6$ on average.  Since the average redshift of the CASSOWARY members is $z\sim0.4$, we expect to see a collection of near-fossil systems, as we are seeing analogs to today's fossil systems in mid-cannibalization of their $L^*$ members.  An interesting thing to note is that if one assumes {\it all} the CASSOWARY progenitors and fossils become/stay fossils, one arrives at a $z=0$ Jones fossil percentage of $36.6\pm2.4$\% and a Dariush percentage of $46.1\pm2.3$\%; this implies we should see far more nearby fossil systems than we currently do.  One explanation why we do not see such an overabundance of nearby fossils is that fossils are transitory in nature, and the look back time is long enough to allow some bright galaxies to fall within half of the fossil's virial radius thereby breaking the fossil's status.  Another explanation is that in the CASSOWARY strong lensing sample, we are seeing a subset of systems that are more likely to be fossil systems; these two hypotheses will be explored further in the non-lensing control sample section.

\begin{center}

\begin{longtable*}{llllp{1.0cm}|c|l|c|p{0.8cm}}
\caption{Fossil Status of All CASSOWARY Members} \\
\hline
Name	& RA & Dec &$N_{200}$&$M_{200}\times10^{14}$M$_{\odot}$&P$_J$/P$_{D}$/F$_{J}$/F$_{D}$& $\Delta m_{12}$/$\Delta m_{14}$ & $\Delta m_{12}^{merge}$/$\Delta m_{14}^{merge}$&$t_{merge}$ (Gyr)\\
\hline
\hline
CSWA 1	& 177.1381\degree& 19.5008\degree&5&0.32& x/x/\checkmark/x & {\bf2.3}/2.4 &&\\
CSWA 2	&159.6816\degree&48.8216\degree&10&0.58& \checkmark/\checkmark/x/x & 0.2/2.4 & {\bf3.3}/{\bf3.7}&0.9\\
CSWA 3	&190.1345\degree&45.1508\degree&12&0.67& x/x/x/x & 0.2/1.4 &  \\
CSWA 4	&135.3432\degree&18.2423\degree&32&1.52& x/x/x/\checkmark & 1.4/{\bf3.1} & &\\
CSWA 5	&191.2126\degree&1.1122\degree&14&0.78& x/\checkmark/x/x & 0.2/1.6 & 1.9/{\bf2.9}&4.1\\
CSWA 6 	&181.5087\degree&51.7082\degree&18&0.94& x/x/\checkmark/\checkmark& {\bf2.2}/{\bf2.8} &&   \\
CSWA 7	&174.4169\degree&49.6099\degree&20&1.05& x/x/x/x & 1.0/1.2 &&   \\
CSWA 8	&182.3487\degree&26.6796\degree&61&2.60& x/x/x/x & 1.6/2.2 && \\
CSWA 9	&186.8281\degree&17.4311\degree&23&1.16& x/x/x/x & 0.7/1.9 &&  \\
CSWA 10	&339.6305\degree&13.3322\degree&30&1.43& x/\checkmark/x/x & 1.4/2.2 & 1.9/{\bf2.8}&3.9\\
CSWA 11	&120.0544\degree&8.2023\degree&26&1.28& x/x/\checkmark/x & {\bf2.0}/2.1&& \\
CSWA 12	&173.3049\degree&50.1445\degree&45&2.00& x/x/x/x & 0.7/1.4 &   \\
CSWA 13	&189.4008\degree&55.5619\degree&26&1.27& x/x/x/x & 0.4/1.8 &   \\
CSWA 14	&260.9007\degree&34.1995\degree&18&0.93&\checkmark/\checkmark/x/x & 0.6/1.9 & {\bf2.5}/{\bf2.8}&3.6\\
CSWA 15	&152.2491\degree&19.6215\degree&51&2.24& x/x/x/x & 0.3/1.6 &&  \\
CSWA 16	&167.7653\degree&53.1486\degree&79&3.20& x/x/x/x & 0.1/1.1 &&  \\
CSWA 17	&174.5373\degree&27.9085\degree&74&3.04& x/x/x/x & 1.3/1.4 &&  \\
CSWA 18	&173.5281\degree&25.5598\degree&25&1.25& x/x/x/x & 1.1/2.0 &&  \\
CSWA 19	&135.0110\degree&22.5680\degree&17&0.89& x/x/x/x & 0.6/0.8 &&  \\
CSWA 20	&220.4548\degree&14.6890\degree&1&0.09& x/x/x/x & && \\
CSWA 21	&5.6705\degree&14.5196\degree&28&1.36& x/x/x/x & 1.0/1.3 &&   \\
CSWA 22	&26.7334\degree&-9.4979\degree&60&2.56& x/x/x/x & 1.1/1.8 &&   \\
CSWA 23	&126.8701\degree&22.5483\degree&88&3.49& x/x/x/x & 1.1/1.5 &&  \\
CSWA 24	&227.8281\degree&47.2279\degree&13&0.71&\checkmark/\checkmark/x/x & 2.0*/2.6* & {\bf2.1}/{\bf2.7}&2.2\\
CSWA 25	&162.4298\degree&44.3432\degree&50&2.21& x/x/x/x & 1.3/1.6 &&  \\
CSWA 26	&168.2944\degree&23.9443\degree&83&3.35&\checkmark/x/x/\checkmark & 1.5/{\bf2.5} & {\bf2.9}/{\bf3.4}&2.1\\
CSWA 27	&247.4773\degree&35.4776\degree&63&2.66& x/x/x/x & 0.2/0.7 &&  \\
CSWA 28	&205.8869\degree&41.9176\degree&31&1.48&\checkmark / \checkmark/x/x &1.8/2.3& {\bf2.1}/{\bf2.8}&1.6\\
CSWA 29	&130.0877\degree&10.8702\degree&2&0.18&  x/x/x/x & &&  \\
CSWA 30	&132.8604\degree&35.9705\degree&31&1.46& x/\checkmark/x/x & 0.7/2.2 &1.1/{\bf2.6}&2.0\\
CSWA 31	&140.3573\degree&18.1715\degree&5&0.32& x/x/\checkmark/x & {\bf2.1**}/{\bf**} &&  \\
CSWA 32	&153.7740\degree&55.5051\degree&1&0.09& x/x/x/x &  &&   \\
CSWA 33	&162.3475\degree&35.7447\degree&26&1.26& x/x/x/x &1.0/2.4 &&  \\
CSWA 34	&4.2564\degree&-10.1531\degree&3&0.19& x/x/x/x & &&  \\
CSWA 35	&149.4133\degree&5.1589\degree&9&0.52& x/x/x/x &1.1/1.3 &&  \\
CSWA 36	&181.8996\degree&52.9165\degree&26&1.25& x/\checkmark/x/x &0.4/1.3 &0.6/{\bf2.6}&3.0\\
CSWA 37	&199.5480\degree&39.7075\degree&19&0.98& \checkmark/\checkmark/x/x &0.8/1.7 &{\bf3.0}/{\bf3.3}&3.5\\
CSWA 38	&186.7154\degree&21.8737\degree&167&5.97& x/x/x/x &1.1/1.4 &&  \\
CSWA 39	&231.9376\degree&6.8761\degree&98&3.83& x/x/x/x &0.3/0.4 &&  \\
CSWA 40	&148.1676\degree&34.5795\degree&77&3.13& x/x/x/x & 0.1/0.3 && \\
CSWA 41	&222.6277\degree&39.1386\degree&20&1.03& x/x/\checkmark/\checkmark& {\bf2.2}/{\bf2.7} &&   \\
CSWA 102 &14.7039\degree&-7.3660\degree&5&0.30& x/x/\checkmark/\checkmark&{\bf2.5}/- && \\
CSWA 103 &26.2679\degree&-4.9311\degree&4&0.27& x/x/x/x  & &&  \\
CSWA 104 &167.5738\degree&64.9965\degree&2&0.12& x/x/x/x & && \\
CSWA 105 &168.7683\degree&16.7606\degree&7&0.41&\checkmark/\checkmark/x/x &0.9/1.8&{\bf2.8}/-&4.7\\
CSWA 107 &176.8471\degree&33.5314\degree&41&1.85& x/x/x/x &0.9/1.5 &&  \\
CSWA 108 &179.0228\degree&19.1868\degree&13&0.70& x/x/x/x &1.1/1.4 &&  \\
CSWA 111 &345.0719\degree&22.2249\degree&25&1.23& x/x/x/x &1.1/1.6 &&  \\
CSWA 116 &25.9589\degree&16.1274\degree&7&0.85&\checkmark/x/x/x &1.9/2.2 &{\bf2.9}/{\bf3.4}&2.0\\
CSWA 117 &150.5106\degree&60.3404\degree&30&1.44& x/\checkmark/x/x &0.8/1.6 &1.9/{\bf2.7}&3.4\\
CSWA 128 &299.6473\degree&59.8495\degree&42&1.90& x/\checkmark/\checkmark/x &{\bf2.0}/2.3 &{\bf2.1}/{\bf2.7}&1.9\\
CSWA 139	&121.8814\degree&44.1800\degree&8&0.45&\checkmark/\checkmark/x/x &0.8/2.0 &{\bf2.6}/-&3.4\\
CSWA 141	&131.6977\degree&4.7680\degree&22&1.11& \checkmark/x/x/x &1.4/1.8 &{\bf2.0}/2.4&0.6\\
CSWA 142	&133.6197\degree&10.1373\degree&54&2.34&\checkmark/\checkmark/x/x &0.6/1.2 &{\bf2.}1/{\bf3.1}&2.9\\ 
CSWA 159	&335.5358\degree&27.7596\degree&6&0.36&\checkmark/x/x/\checkmark&1.3/{\bf2.8} &-/-&3.6\\
CSWA 163	&329.6820\degree&2.9584\degree&18&0.93& x/\checkmark/x/x &1.4/2.1 &1.8/{\bf2.7}&3.2\\
CSWA 164	&38.2078\degree&-3.3906\degree&10&0.55& x/x/x/\checkmark &1.7/{\bf3.2} &&  \\
CSWA 165	&16.3318\degree&1.7489\degree&14&0.78& x/x/x/\checkmark &1.7/{\bf2.8}  &&  \\
\hline
\caption{\footnotesize{All 58 CASSOWARY members and their general properties.  Since both Jones et al.\ (2003) and Dariush et al.\ (2010) criteria were used to determine fossil status, we include the magnitude gap in the $r$-band between the first and second rank galaxy ($\Delta m_{12}$) along with the first and fourth rank galaxy ($\Delta m_{14}$)  within $0.5R_{200}$.  The P$_{J/D}$ and F$_{J/D}$ columns were added to differentiate between Jones(J)/Dariush(D) progenitors or fossils, respectively.  Bolded entries indicate optical fossil status being reached either now or after merging is completed.  Dashes under the merged column indicate all galaxies within $0.5R_{200}$ merging into one BGG by $z=0$.  $t_{merge}$ indicates the expected merger time scale from Kitzbichler \& White's (2008) relation until fossil status in acheved.  *\;Double nucleus detected in archival {\it HST} imaging negating fossil status until merging is finished.  **$\;${\it Gemini} GMOS data from Grillo et al.\ (2013).}}
\label{fig:data}
\end{longtable*}
\end{center}

\begin{figure*}
\vspace{-1.2truein}
\hspace{-0.20truein}
\begin{tabular}{cc}
	{\includegraphics[scale=0.44]{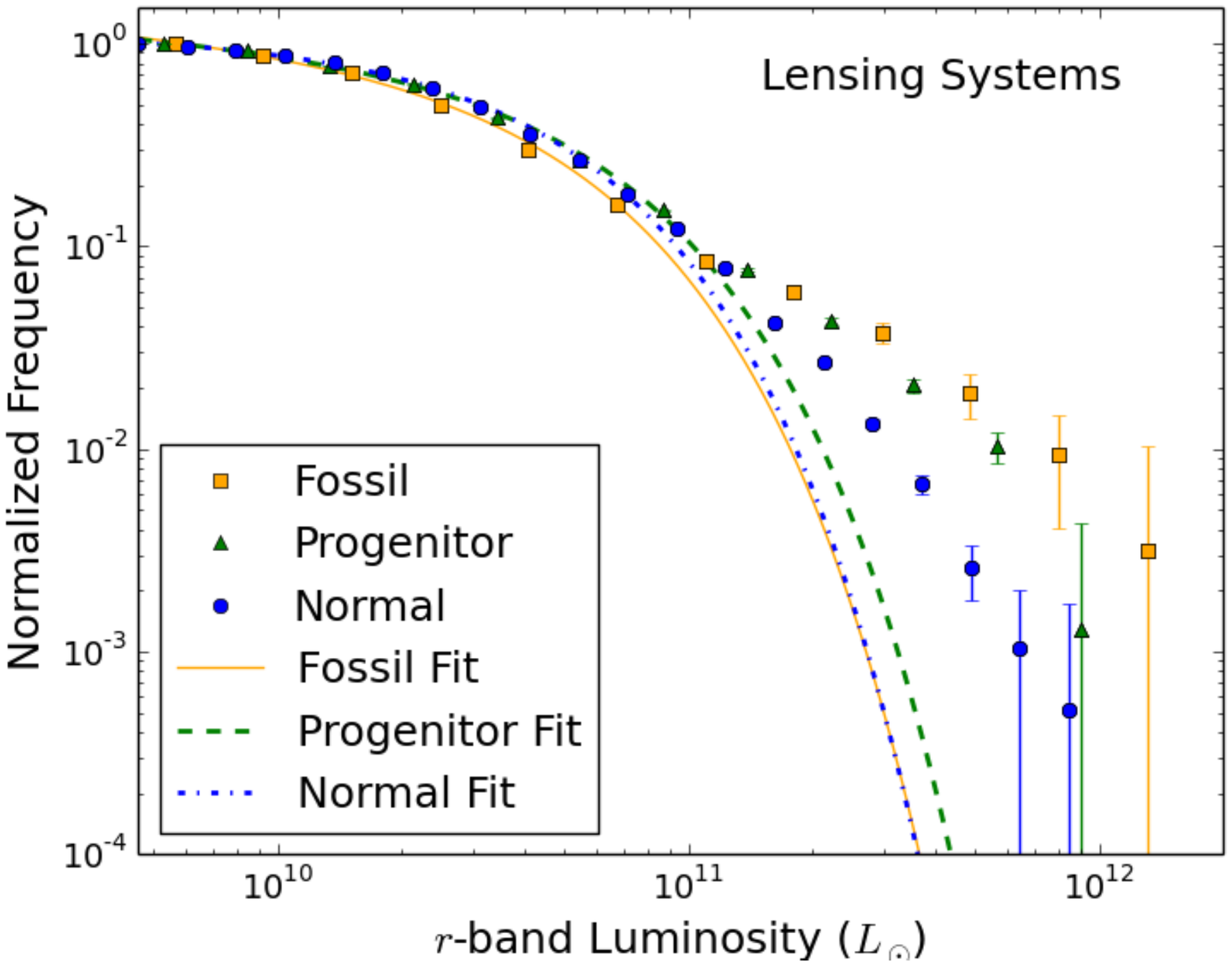}}&\hspace{-0.45 truein}{\includegraphics[scale=0.44]{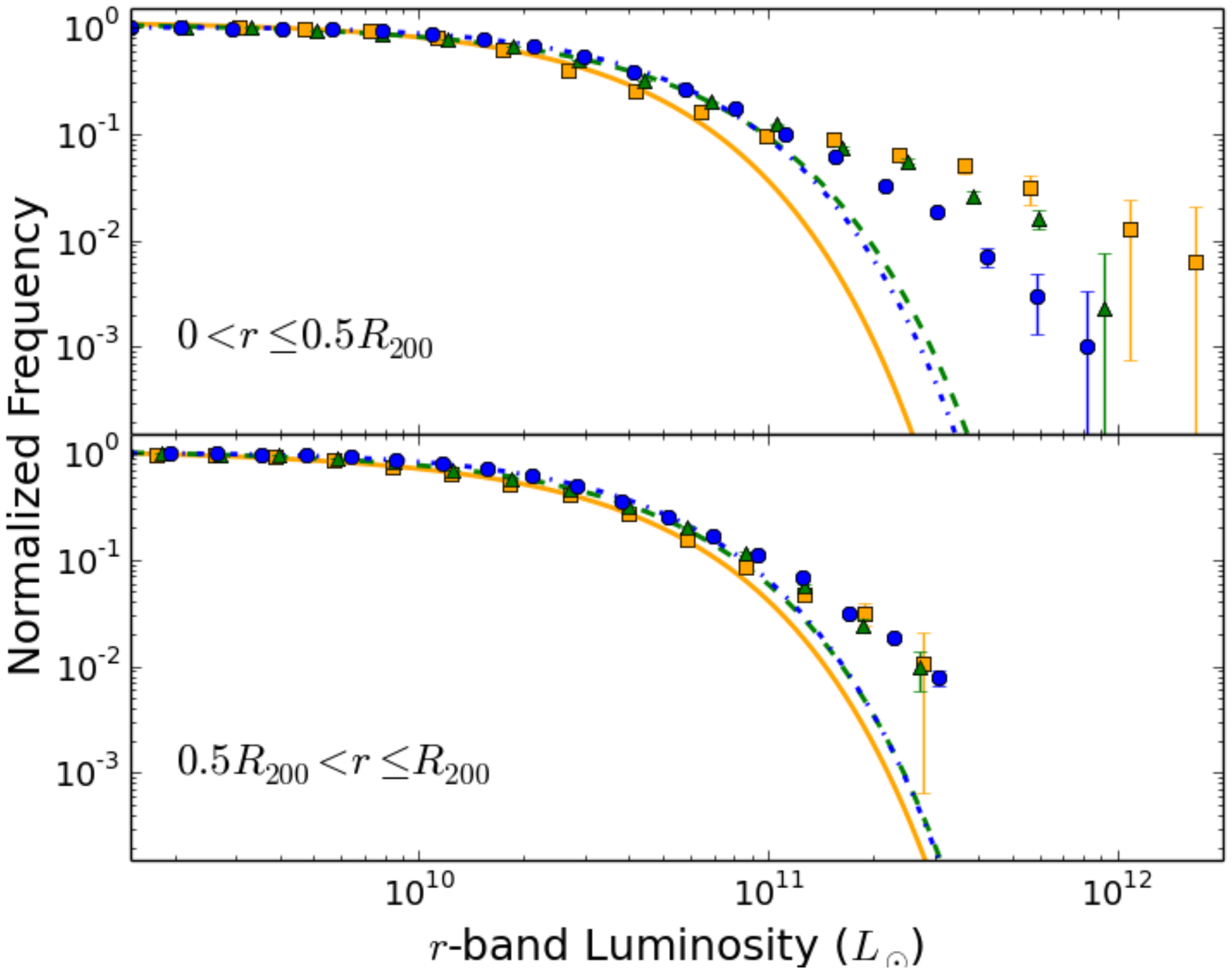}} \\
\end{tabular}
\vspace{-1.0truein}
\caption{\footnotesize
{{\it Left:} Cumulative luminosity functions of CASSOWARY systems separated by classification: normal, progenitor, and fossil.  The curves represent the best fit Schechter function found using least squares fitting for each population excluding the BGGs.  The progenitor function falls between the normal and fossil functions at the bright end ($L\gtrsim 10^{11}$ L$_{\odot}$), supporting the idea that fossil progenitors are a bridge between normal and fossil systems.  {\it Right:} The same luminosity functions binned by inner and outer half virial radii.  The inner regions exaggerate differences between the three populations which may be due to frequent mergers/interactions of members.  The outer regions show little statistical difference from one another, however as a whole, the outer members possess more bright galaxies than the best-fit model would suggest.}  Error bars are at $1\sigma$.}
\label{fig:lensing}
\end{figure*}

To better determine if we are truly seeing the progenitors of today's fossil systems, we contrasted the galaxy luminosity functions of each category against one another, as fossil system luminosity functions show a clear deficit in $L^*$ members when compared to comparable sized normal groups and clusters (Gozaliasl et al.\ 2014).  Fossil progenitors might be expected to be a transitional step between the two extremes, losing $L^*$ galaxies as they are consumed by the BCG's.  We created three average cumulative galaxy luminosity functions (a fossil, progenitor, and normal function) from the CASSOWARY lensing sample to ensure we were comparing strong lensing systems to other strong lensing systems{\footnote[4]{In the cases of split identification (e.g., Jones fossil and Dariush progenitor), the Jones criterion was chosen for forming the luminosity functions since it is the most widely cited.}}.  Due to the large amount of overlap between the two prevailing fossil criteria (Jones/Dariush) in this sample, the cumulative luminosity functions combined both Jones and Dariush fossils/progenitors to minimize errors. It is important to note that the poor fitting at the bright end is due to the ``BCG bump," a known artifact of galaxy mergers in the centers of clusters and groups (Hansen et al.\ 2005).  Excluding the BGGs, we found that overall, the lensing fossil and normal population fits are nearly identical.  However, when the BGGs are introduced into the data set, we found that the fossil luminosity function greatly diverged from the normal luminosity function at the bright end, as expected (Figure~\ref{fig:lensing}). The progenitor luminosity function matched the normal function at the faint end.  However, the progenitor function fell between the normal and fossil functions at the bright end suggesting that fossil progenitors are currently losing their intermediate members while gaining very bright members, thereby moving the groups closer to fossil status.  This supports the notion that fossil systems form their massive BCGs via cannibalization of intermediate mass member galaxies.  SDSS images of the inner regions of groups classified as fossil progenitors also very often show an extremely crowded environment near the BCG further supporting this mechanism of fossil formation (Figure 6). 


Much work has been done investigating the global deficit in intermediate-luminosity members and the value/evolution of the faint-end slope of the fossil luminosity function, finding that the deficit in $L^*$ members is likely due to cannibalization by the BCG and the faint end slope is consistent with normal groups (Lieder et al.\ 2013, Gozaliasl et al.\ 2014, Zarattini et al.\ 2015).  However, little work has gone toward investigating fossil populations in different radial bins, where initial group conditions could still be encoded (particularly in the outer regions).   Binning the average luminosity functions into inner ($r\leq0.5R_{200}$) and outer ($0.5R_{200}<r\leq R_{200}$) regions reveals that this deficit in intermediate mass galaxies/abundance of extremely bright galaxies in fossil progenitors and fossil systems is exaggerated for $r\leq0.5R_{200}$ (Figure~\ref{fig:lensing}; right).  This is likely due to the increased galaxy density near the center effectively speeding up the galaxy interaction rate,  
therefore the central regions of fossil or near-fossil systems should show the most extreme differences from non-fossils.  While the inner progenitor fit is nearly identical to the normal fit, when the BGGs are included in the histogram a clear difference can be seen, placing it firmly between the normal and fossil functions.

We quantified the statistical significance of these differences between datasets via a one-sided K-S test which gives the probability that differing data sets come from the same cumulative distribution function.  For $0<r\leq 0.5R_{200}$, even with the relative lack of data points, lensing fossil systems showed only a 0.84\% chance of being identical to normal lensing systems.  Due to a larger sample size, lensing progenitors also proved to be significantly different than normal lensing systems with only a 0.01\% chance of being identical within $r\leq 0.5R_{200}$.  Unfortunately, due to insufficient galaxy counts in the lensing fossil population, we were not able to find a significant difference between lensing fossils and progenitors.  The outer half virial radii galaxies showed no statistically significant differences between the populations, suggesting that most differences in galaxy populations for fossil systems exist near the center where the most processing has occurred.  However, all lensing systems exhibit an average $\sim2\sigma$ deviation from a Schechter function in the outer $0.5R_{200}$ for galaxies brighter than $L^*$ which is surprising.  Since these galaxies are farther away from the center, one would expect them to be much less processed and therefore be better represented by the fits.  Additionally, there are no BGGs in the outer regions to skew the data away from a Schechter function.  While by no means definitive, this suggests that lensing systems may form differently from non-lensing systems of comparable mass.

\smallskip
\subsection{\textbf{Non-lensing Control Sample}}

\begin{figure*}
\vspace{-1.0truein}
\hspace{-0.20truein}
\begin{tabular}{cc}
	{\includegraphics[scale=0.44]{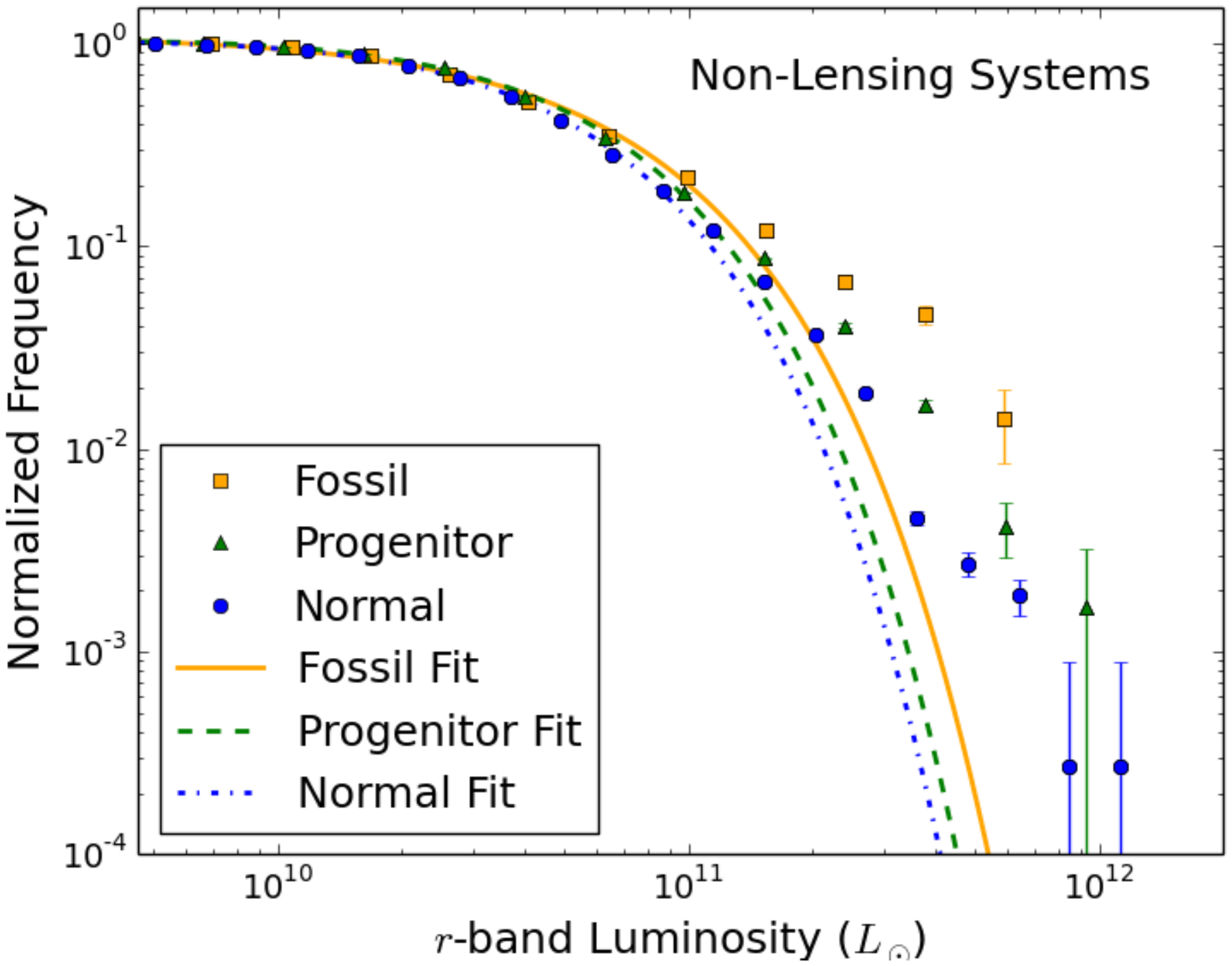}}&\hspace{-0.45 truein}{\includegraphics[scale=0.44]{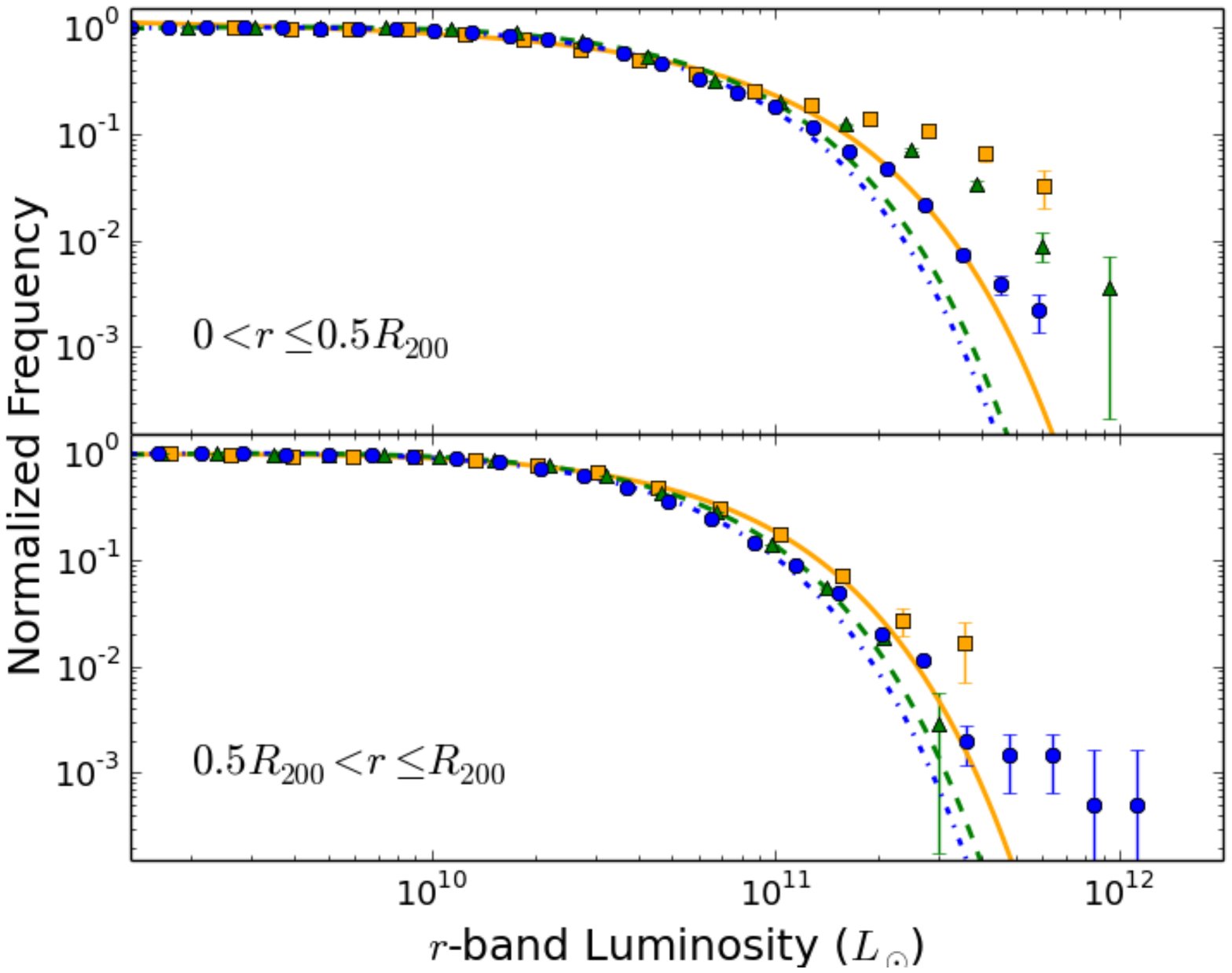}} \\
\end{tabular}
\vspace{-1.0truein}
\caption{\footnotesize
{{\it Left}: Cumulative luminosity functions of our non-lensing control sample for all member galaxies.  For non-lensing groups, differences in the bright end (while still visually apparent) are not as prominent as our lensing systems, supporting the existence of a strong lensing bias.  {\it Right}: The same non-lensing luminosity functions binned by inner and outer half virial radii.  The inner regions for non-lensing fossils and progenitors also show the excess of bright member galaxies when compared to normal groups, though again, less pronounced than in our lensing sample.  The outer regions show virtually no difference between progenitors and normal groups, with fossils only housing a few more bright galaxies.  Error bars are at $1\sigma$.}}
\label{fig:control}
\end{figure*}

All members of our sample exhibit strong gravitational arcs near the BGGs, implying a high central mass concentration for these systems.  To see how/if the presence of strong gravitational arcs biases our fossil system and progenitor findings, we assembled a one-to-one random control sample of non-lensing groups from the Augmented maxBCG cluster catalog (Rykoff et al.\ 2012), Clusters of galaxies in SDSS-III (Wen et al. 2012), and Richness of Galaxy Clusters (Oguri 2014) catalogs.  Control groups were selected to match (within 10\%) each CASSOWARY group in both redshift and galaxy richness, and when multiple matches for control groups were found among the catalogs, the closest match was chosen.  To increase the accuracy of this one-to-one comparison, we found two non-lensing matches for each CASSOWARY group\footnote[5]{Only one non-lensing match was able to be found for CSWA 31 due to its high redshift ($z=0.683$) and poor member count ($N_{200}=5$).}.  Using the same photoZ and color cuts as the lensing sample, we found fossil percentages of $2.9\pm1.6$\% (Jones) and $13.6\pm1.2$\% (Dariush) and fossil progenitor percentages of $17.5\pm1.2$\% (Jones) and $25.2\pm1.1$\% (Dariush) showing that while being a strong gravitational lens does not significantly alter the Dariush fossil fraction, it does greatly improve the chances that a fossil will be a classic Jones fossil.  The progenitor fraction is consistent between the control sample and lensing sample suggesting that in general the progenitor fraction is not greatly affected by the presence of gravitational arcs.  It is important to note that while the non-lensing Dariush fossil fraction is consistent with the prediction from Gozaliasl et al.\ (2014) for $z\leq0.6$, our non-lensing Jones fossil fraction of $2.9\pm1.6$\% is far below their estimate of $22\pm6\%$.  This could be partially due to the Gozaliasl et al.\ (2014) $\Delta m_{12}\geq1.7$ being less restrictive than the classical Jones et al.\ (2003) $\Delta m_{12}\geq2.0$.  Also, while their sample includes all groups within $z\leq0.6$, we only have two groups within $z\leq0.2$ making our sample somewhat different from theirs.


To see if the lower fossil occurrence rate in our sample could be due to inadvertantly including too many bright galaxies (since spectroscopic data is lacking for these groups) we calculated the bright galaxy ($L_{\odot}>0.4L^*$) overdensity within each CASSOWARY group compared to the surrounding regions.  In non-fossil CASSOWARY groups, an overdensity of bright galaxies (sufficient to prevent that group from being a fossil) within $0.5R_{200}$ was confirmed above $13\sigma$ confidence, indicating our fossil fractions are reliable.  The apparent fossil deficit could be accounted for due to the redshift range of our sample ($0.2<z<0.7$).  Since most fossils discovered lie near $z\sim0.1$, and Kanagusuku et al.\ (2016) found in the Millennium Simulation that most $z=0$ fossils made the transition between $0.3<z<0.6$, there could be a lower fossil fraction in our samples.

Average galaxy luminosity functions were also generated for the non-lensing sample to compare against the lensing sample to see how strong lensing might affect a group's galaxy population (Figure~\ref{fig:control}). The non-lensing fossil and normal luminosity functions exhibit similar behavior as the lensing sample.  Non-lensing fossils show a 0.01\% chance of being identical to non-lensing normal {\it and} progenitor systems, confirming that on average fossil systems' inner regions house a different population of galaxies than non-fossils.  Interestingly, non-lensing progenitors showed virtually no differences from non-lensing normal systems at any radii; this reinforces our hypothesis that the presence of a strong gravitational arc marks the most extreme examples of fossil formation at all stages.  Also, the non-lensing progenitor fit falls between the other two in each radial bin.  Reintroducing the BGGs maintains this in-between state for the non-lensing progenitors.  Applying the K-S test to the non-lensing populations showed again that significant differences only appear within $0<r\leq 0.5R_{200}$.

\subsection{\textbf{Comparing Lensing vs. Non-Lensing Samples}}

\begin{figure*}
\vspace{-1.0truein}
\hspace{-0.60truein}
\begin{tabular}{cc}
	{\includegraphics[scale=0.49]{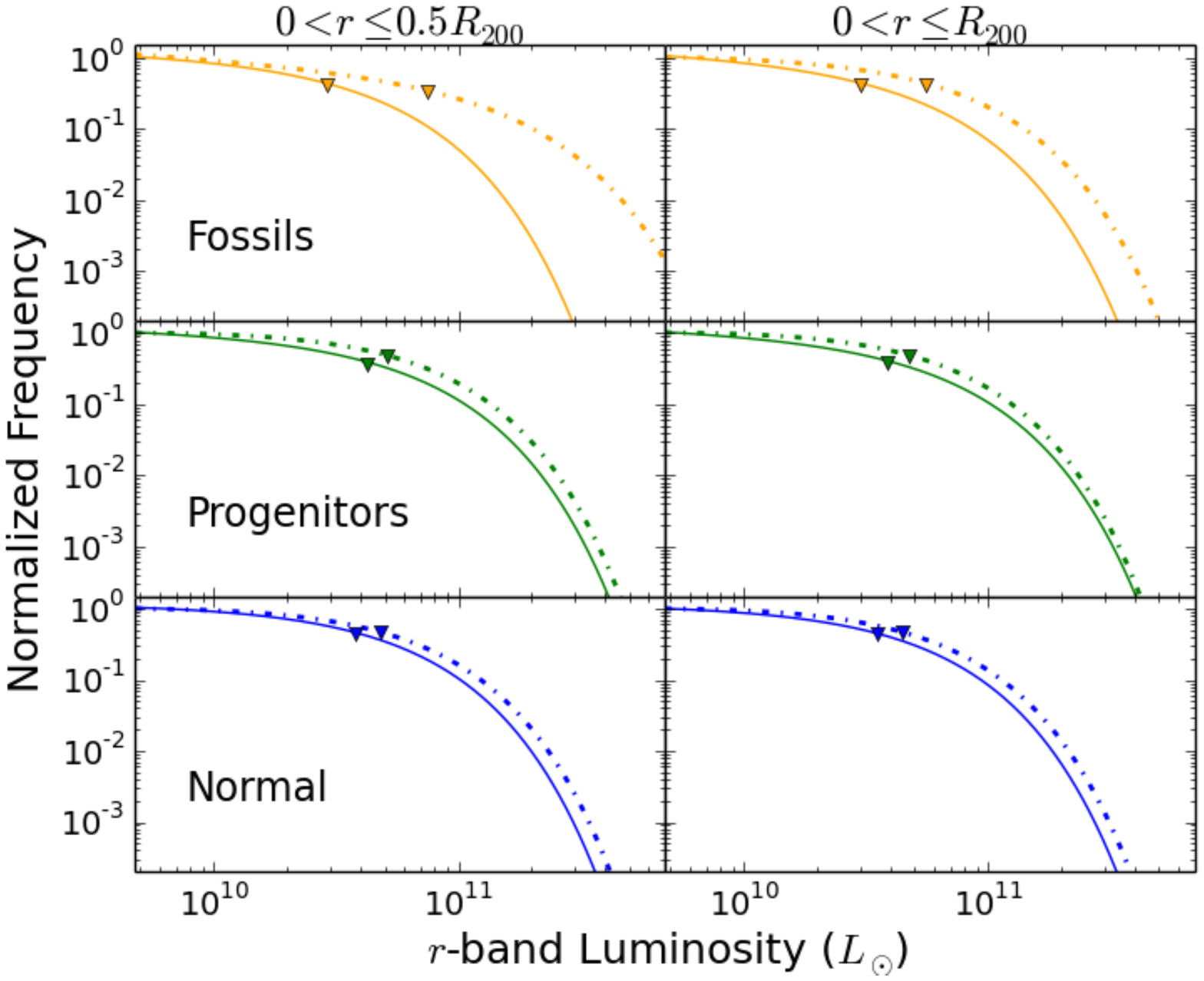}}&\hspace{-0.6 truein}{\includegraphics[scale=0.49]{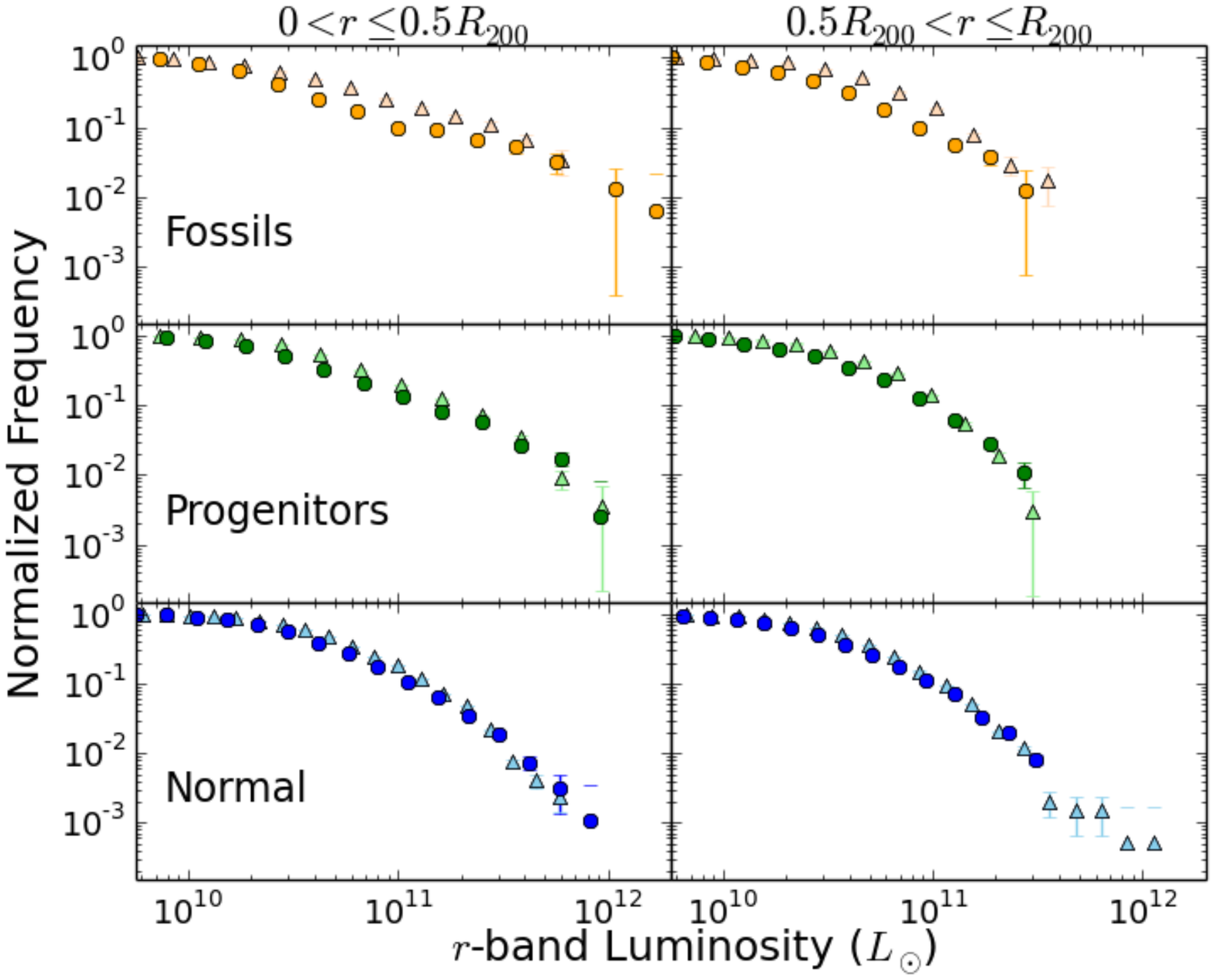}} \\
\end{tabular}
\vspace{-1.0truein}
\caption{\footnotesize
{{\it Left:} Contrasting the best fit Schechter functions, excluding the BGGs, of lensing/non-lensing samples at inner-half and full virial radii bins separated by fossil status.  The CSWA strong lensing fit is shown with a solid line and the non-lensing control fit is represented by the dashed line with the derived $L^*$ for each model marked.  {\it Right:}  Galaxy member data taken from SDSS with BGGs included.  Circles represent the lensing systems, and triangles mark the non-lensing systems.  In each case, lensing groups show a deficit of intermediate-luminosity members and an excess in bright members implying that lensing systems could be an example of the most extreme fossil systems along with them being more likely to become fossils as opposed to similarly sized non-lensing systems. }}
\label{fig:compare}
\end{figure*}

A comparison of the lensed vs.\ non-lensed fossil luminosity function revealed interesting distinctions; on average, lensing fossil systems lack intermediate mass galaxies and house larger BGGs than their non-lensing counterparts (Figure~\ref{fig:compare}).  While the latter is not terribly surprising (as larger galaxies are more likely to act as good gravitational lenses), the former offers help explaining why the lensing sample has significantly more Jones fossils than the non-lensing sample; the lensing sample is the extreme case in fossil formation.  In CASSOWARY fossils (and progenitors to a lesser extent), we are seeing elevated $L^*$ cannibalization resulting in intermediate galaxy deficits and an overrepresentation of extremely bright galaxies.  Comparing the non-lensing progenitor fit to its lensing counterpart reveals even sharper differences between non-lensing fossils and lensing fossils (top of Figure~\ref{fig:compare}). On average, the lensing progenitors have fewer bright $L^*$ galaxies than the non-lensing progenitors.  This could be an indication of a strong lensing selection bias. Since the presence of strong lensing indicates a high mass concentration, lensing progenitors could have already had most of their $L^*$ members consumed by the BGGs.


Since Dariush fossils do not need such a large luminosity difference between the BGG and the next ranked galaxies, the non-lensing control sample holds many more Dariush fossils than Jones fossils.  
The non-lensing control sample also indicates that while the presence of a strong gravitational arc does not strongly affect the likelihood of finding the progenitors to today's fossils around $z\sim0.4$, it does appear to greatly increase the probability of locating Jones fossil systems.  Applying the K-S test, this time to lensing vs.\ non-lensing systems, again shows significant differences only within $0.5R_{200}$.  For the inner regions, normal systems proved to be consistent between lensing and non-lensing systems.  Progenitors, on the other hand, showed a 0.01\% chance of being identical; such a striking result means that a strong lensing bias may very well exist.  Since lensing progenitors on average show different galaxy populations than non-lensing progenitors, it can be inferred that the same (if not more) can be said for lensing vs.\ non-lensing fossils.  Unfortunately, errors in bin count for both fossil populations kept us from arriving at any meaningful result; this can be remedied by increasing the sample size.

When one compares the best fit Schechter functions of lensing vs.\ non-lensing systems as a whole, an interesting distinction is found: at every radial bin, systems acting as strong gravitational lenses exhibit an intermediate-luminosity member deficit regardless of fossil status.  While this is expected near the center of most groups, (the act of forming the BGG consumes many of these galaxies thereby shifting $L^*$ toward the faint end) this deficit supports the existence of a strong lensing bias toward fossil-like systems.  To test whether or not these lensing systems are preferentially selecting systems with different initial conditions from normal groups, thereby supporting the idea that some fossils are born differently than most systems, the outer regions must be probed to see if the galaxy populations differ there as well.  However the overall lack of members in the outer half virial radii of the systems made any conclusions statistically insignificant.

\section{\textbf{Progenitor - Fossil Properties, Timelines, and the Longevity of Fossil Systems}}

\begin{figure}[h]
\vspace{-1.25truein}
\hspace{-0.25truein}
{\includegraphics[scale=0.445]{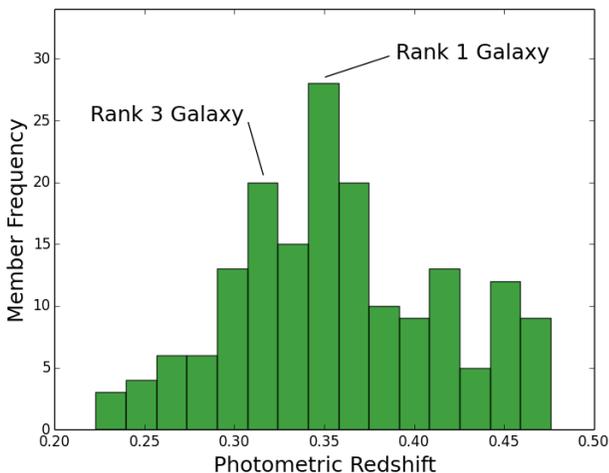}}
\vspace{-1.25truein}
\caption{\footnotesize
{Histogram of all candidate members of the fossil progenitor CSWA 26.  The double peak near $z=0.34$ is suggestive of another group merger scenario similar to the Cheshire Cat (CSWA$\;2$) but on a much larger scale.}}
\vspace{-0.1truein}
\label{fig:CSWA_26_hist}
\end{figure}

Since fossil systems are traditionally believed to be relaxed systems, it stands to reason that fossil IGMs should possess well-developed cool cores.  While some do{\footnote[6]{Although some fossil systems show cool cores, many are smaller than one would expect from the cooling time.}}, there are many which instead show flat radial temperature profiles or even temperature spikes near their centers (Sun et al.\ 2004; Khosroshahi et al.\ 2004, 2006).  More recently, Irwin et al.\ (2015) discovered that the Cheshire Cat fossil progenitor was likely being shock heated near its center because of a group merger, meaning that even after the BGG merger is complete it will be seen as a non-cool core fossil system for $\sim$2 Gyr.  Miller et al.\ (2012) also noticed odd asymmetries in the IGM of many nearby fossils, suggesting that their histories were complex and possibly violent as well.  Such irregularities in the hot gas could be better understood if similar irregularites are also seen in fossil progenitors. Following this thought, one of the most promising systems in the CASSOWARY catalog is CSWA 26. The $z=0.336$ fossil progenitor is six times more massive than the Cheshire Cat and shows some evidence of being another group merger, the result of which would be the most massive fossil cluster yet discovered which does not possess a cool core.  The photometric redshifts of CSWA 26's members indicate a possible double peak about the BCG and the third rank galaxy (Figure \ref{fig:CSWA_26_hist}).  The mean redshift of each peak shows a radial velocity difference of $\Delta v>1000$ km s$^{-1}$ giving weight to this being a large group merger.  

\begin{figure*}
\centering
\begin{tabular}{lccc}
	{a. {\includegraphics[angle=-90, scale=0.1325]{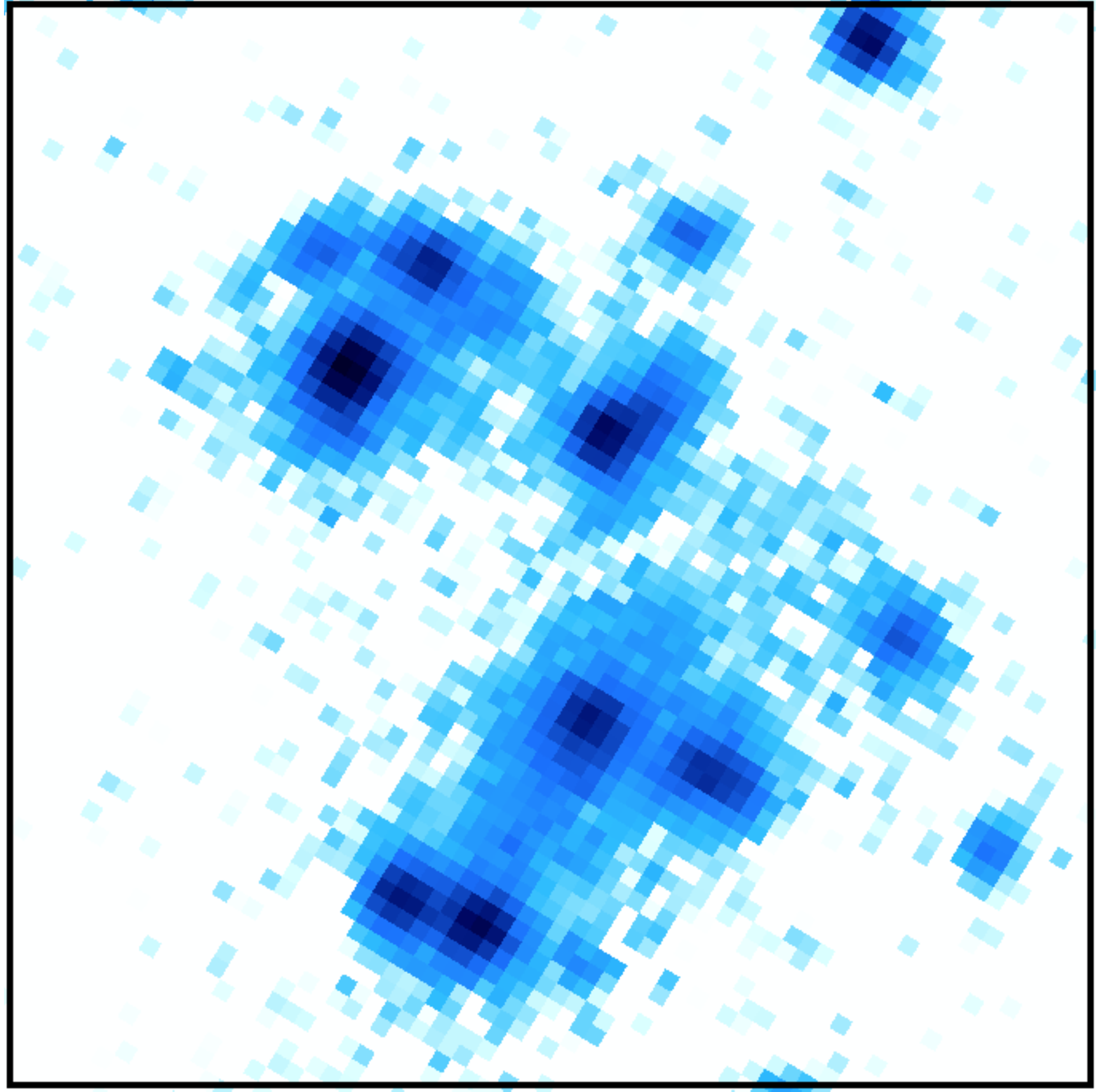}}}&{b. \includegraphics[angle=-90, scale=0.1325]{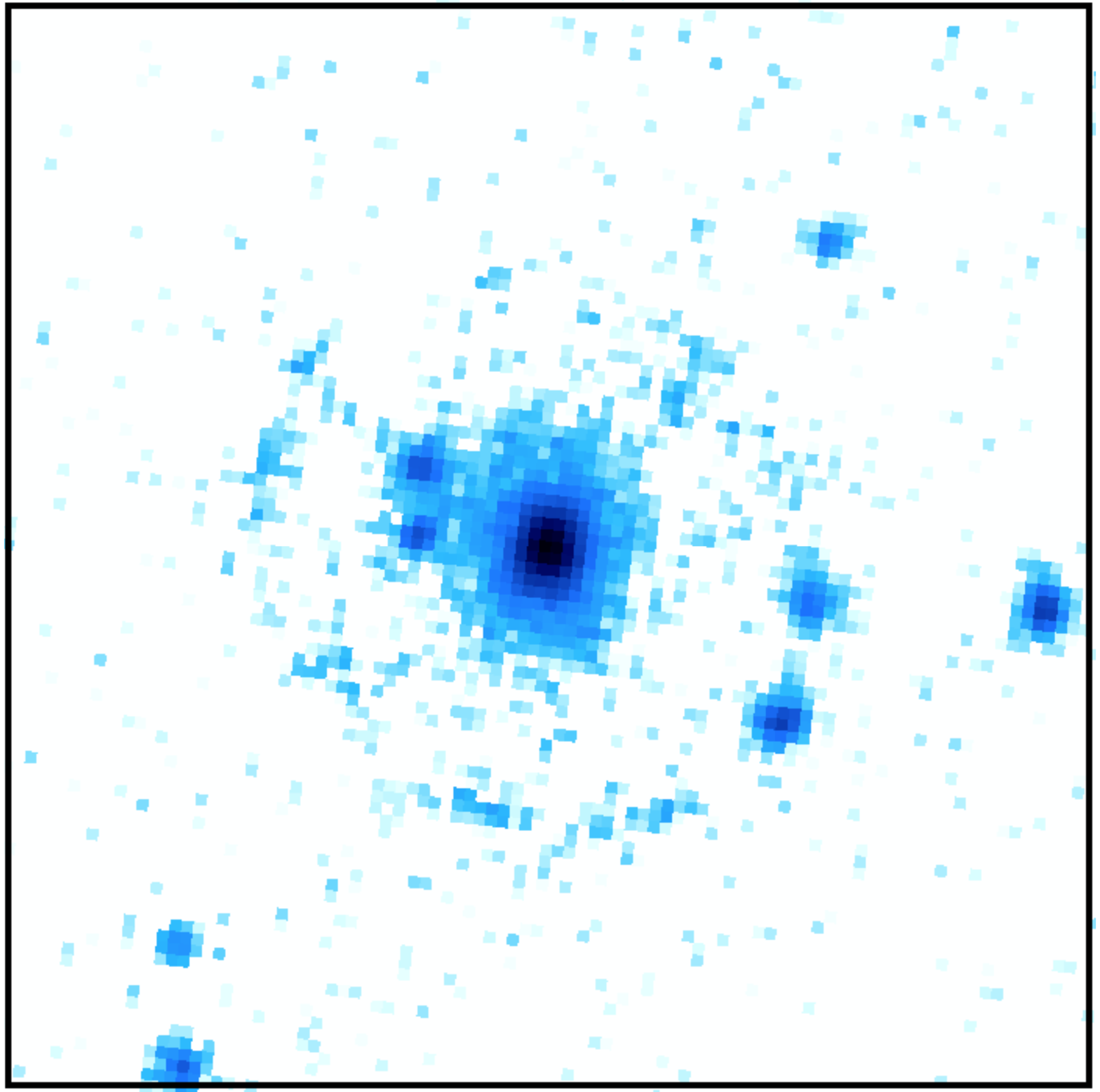}}&{c. \includegraphics[angle=-90, scale=0.1325]{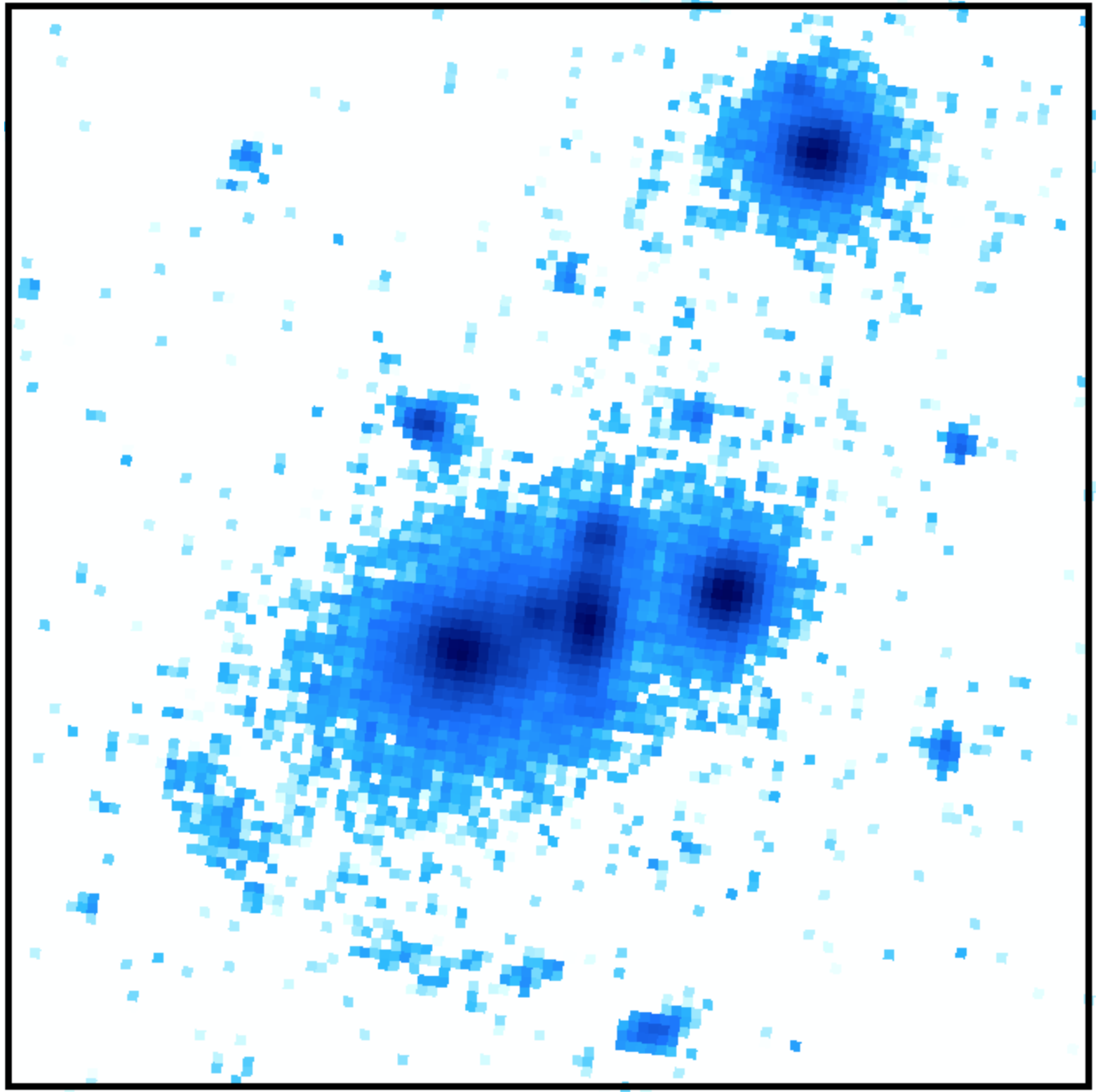}}&{d. \includegraphics[angle=-90, scale=0.1325]{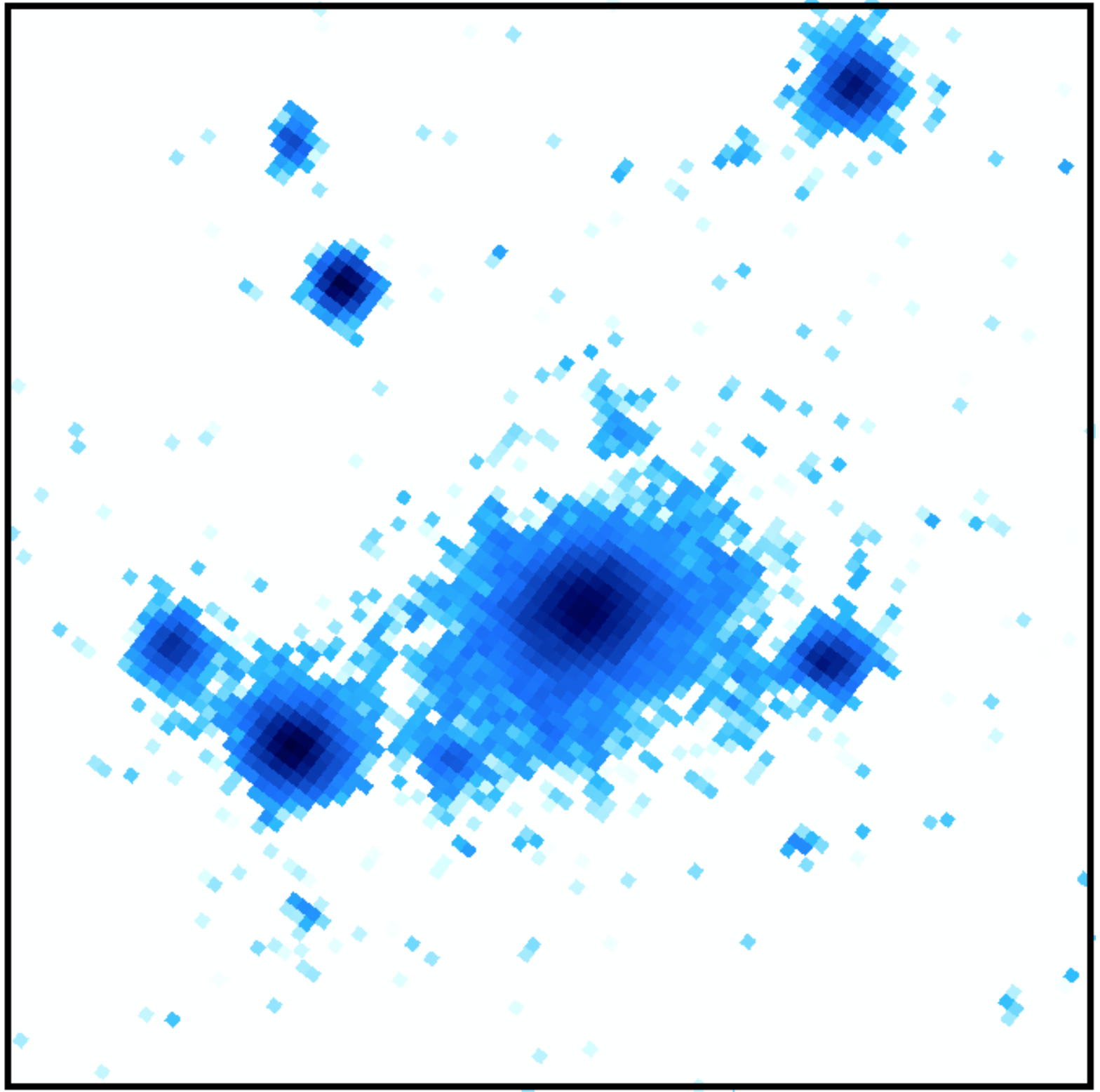}}\\
	{e. \includegraphics[angle=-90, scale=0.1325]{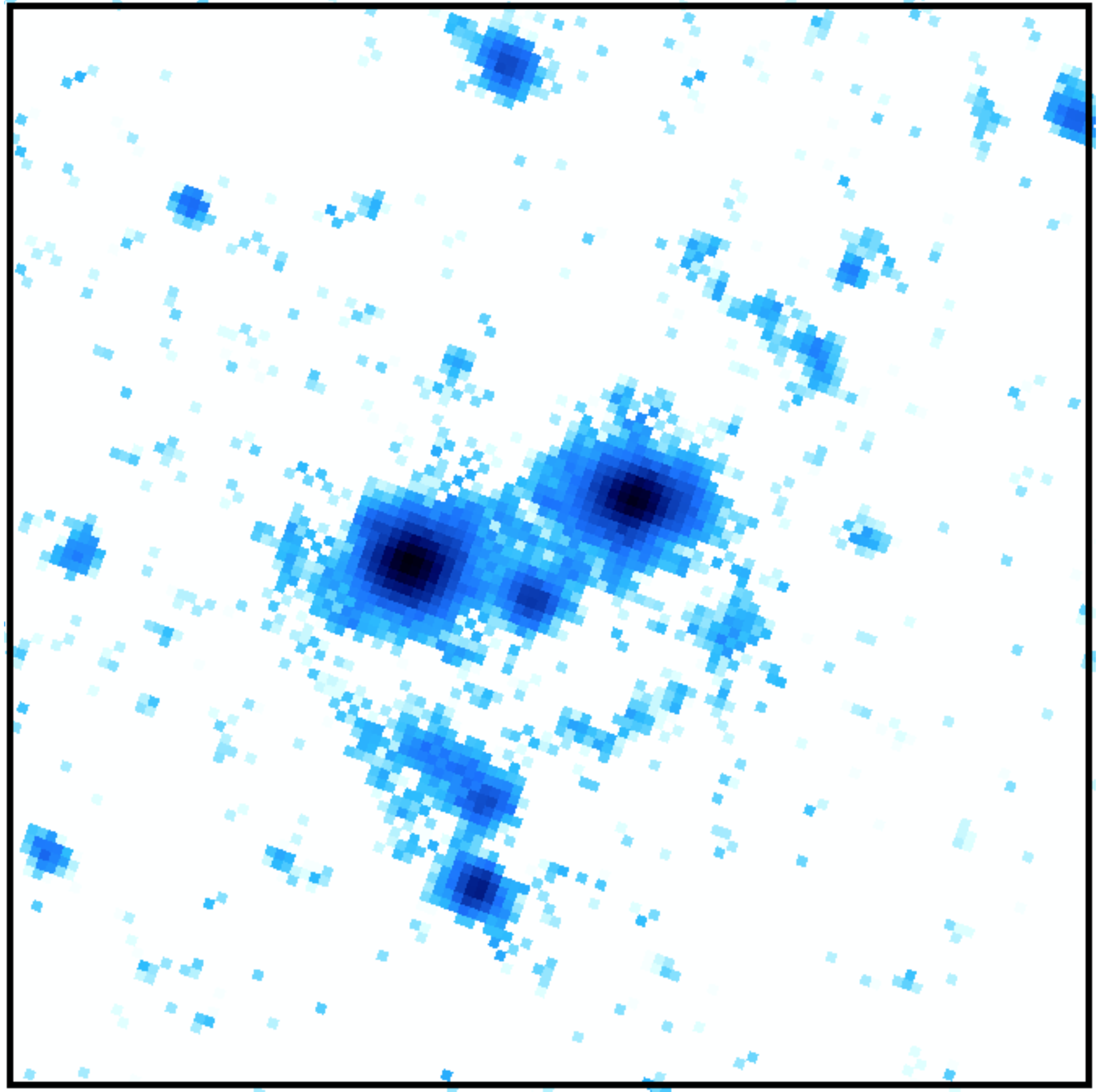}}&{f. \includegraphics[angle=-90, scale=0.1325]{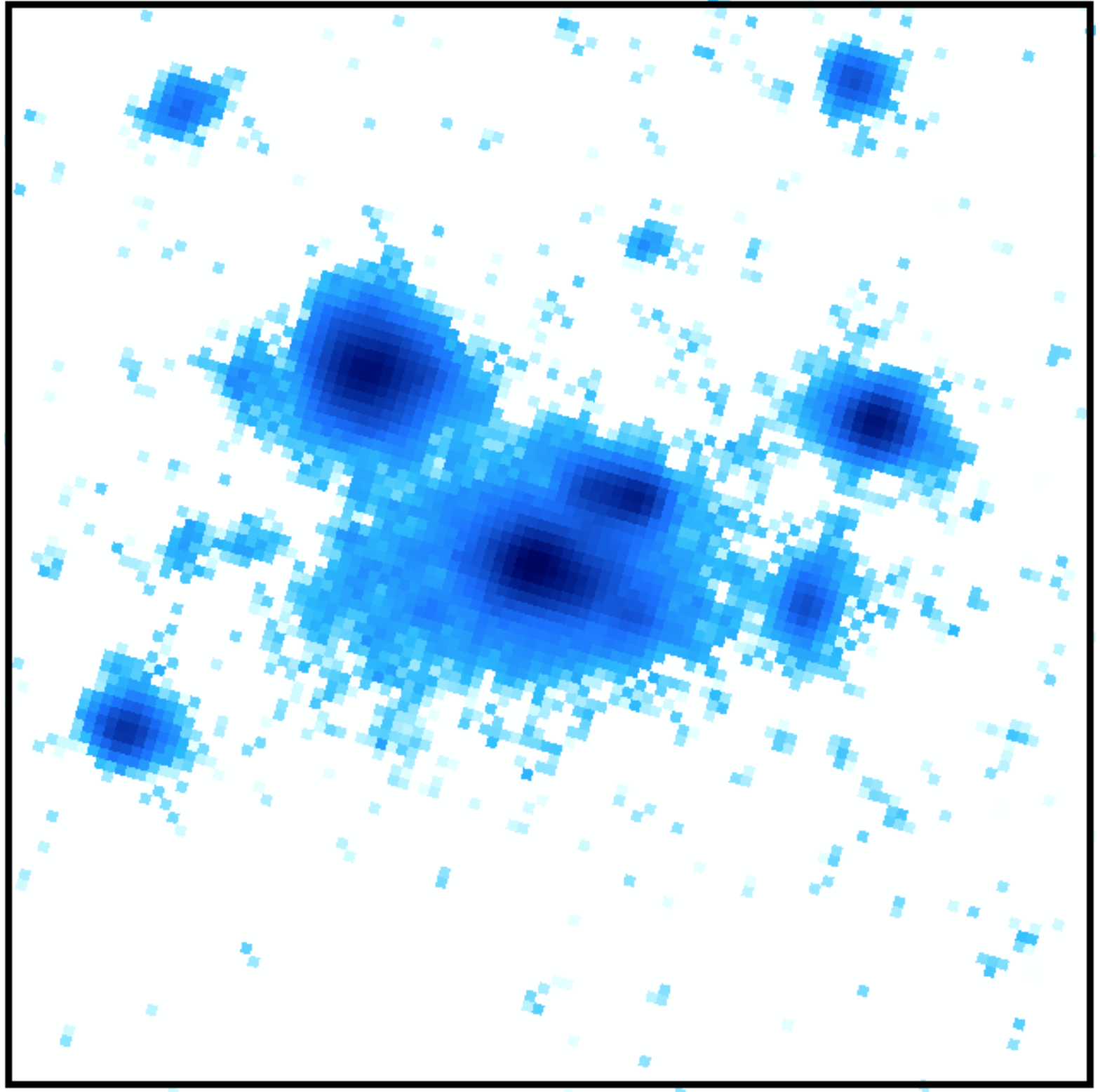}}&{g. \includegraphics[angle=-90, scale=0.1325]{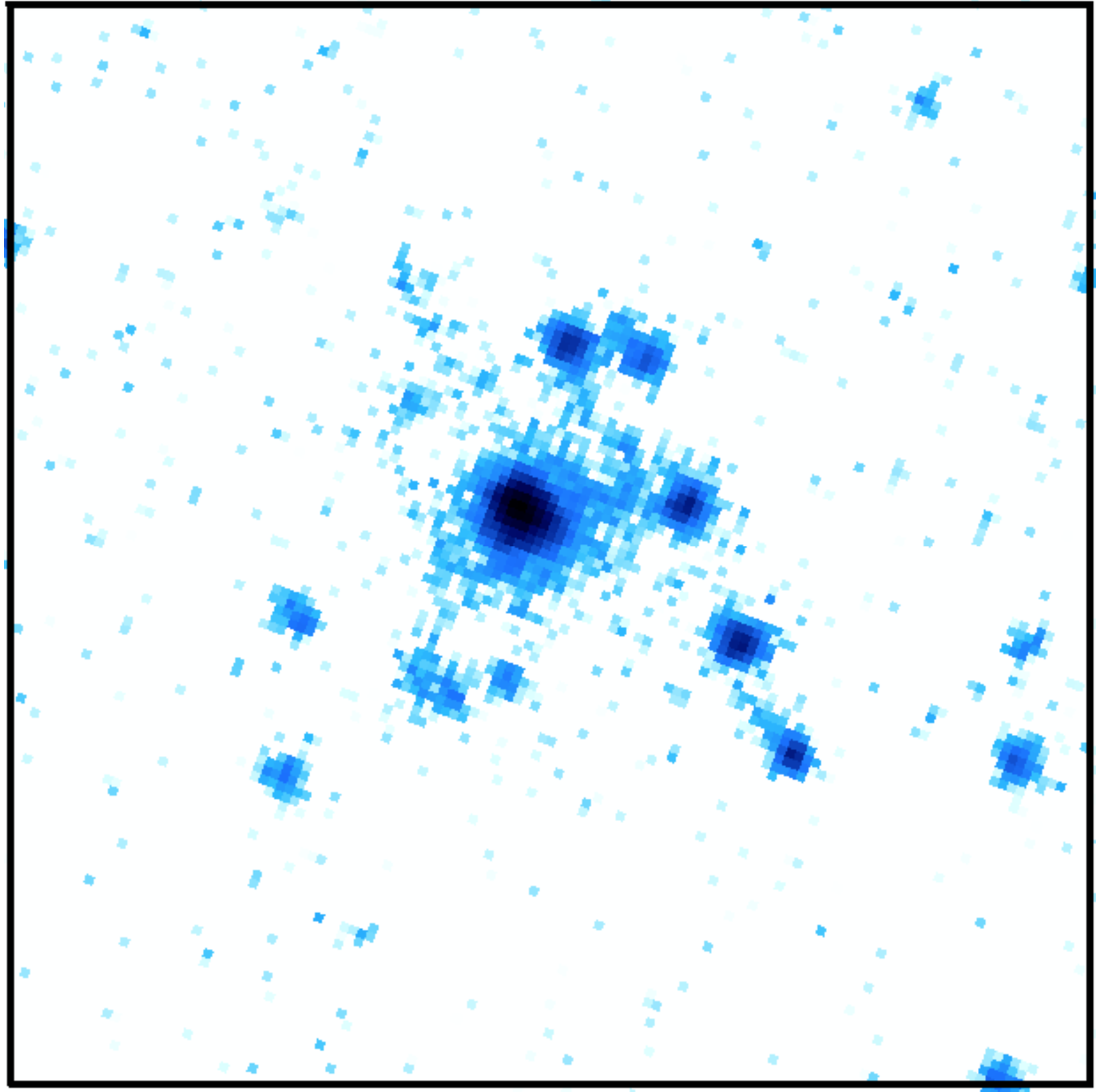}}&{h. \includegraphics[angle=-90, scale=0.1325]{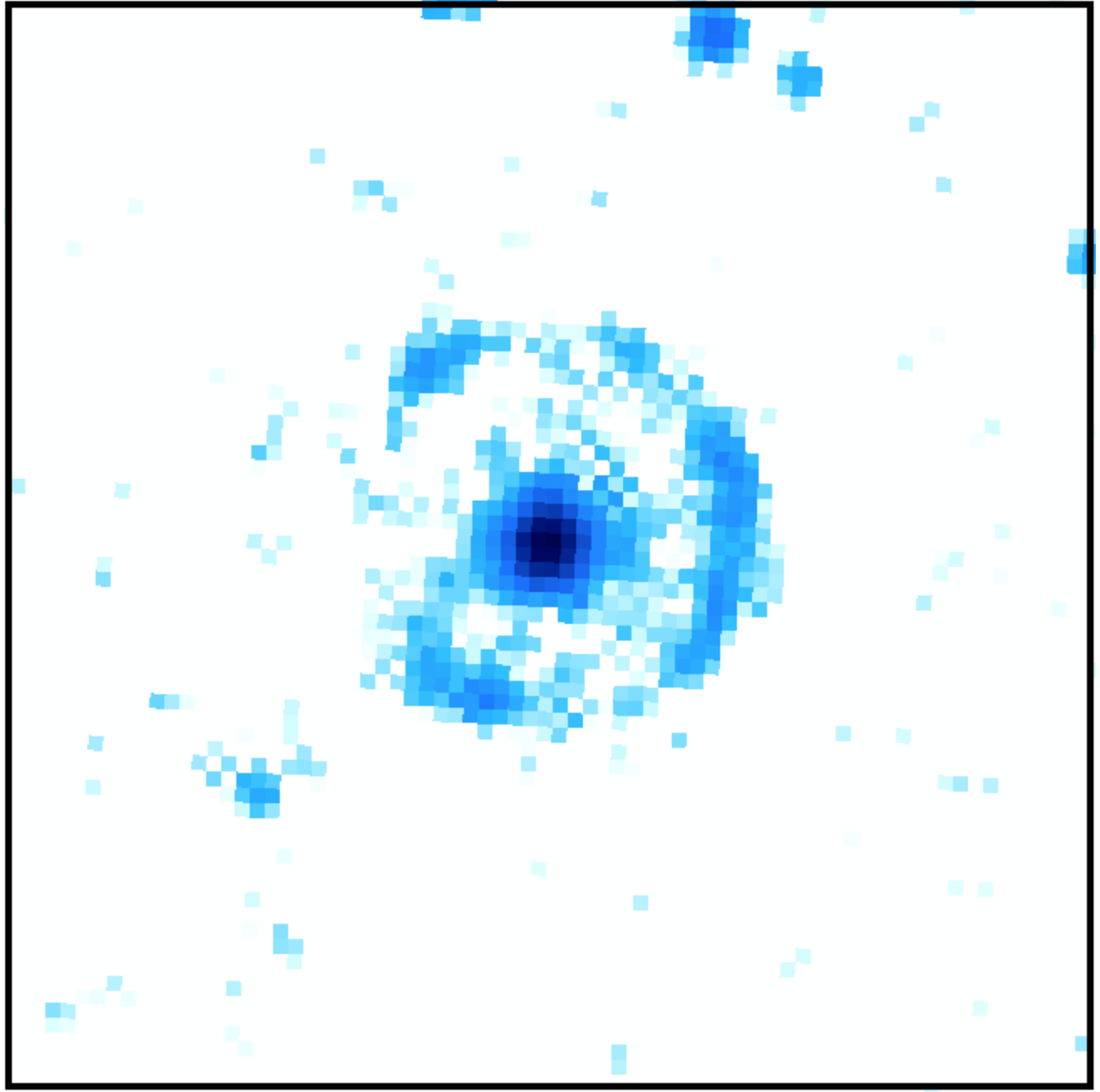}} \\
\end{tabular}
\caption{\footnotesize
{An SDSS collage of CASSOWARY systems at various stages in BGG formation, in time order from a to h.  Appearing in order from upper left to bottom right along with time until fossil formation is complete: normal group (a.) CSWA 23, (b.) CSWA 10 (3.9 Gyr), (c.) CSWA 26 (2.4 Gyr), (d.) CSWA 30 (2.0 Gyr), (e.) CSWA 2 (0.9 Gyr), (f.) CSWA 11 (double nuclei; $<100$ Myr), (g.) CSWA 4 (young fossil with ongoing mergers), (h.) CSWA 1 (relaxed fossil).}}
\label{fig:collage}
\centering
\end{figure*}

Since we have many progenitors in the CASSOWARY catalog with a wide range of merging time scales until transitioning into a fossil system (Table 2), we can form a rough timeline of today's average fossil system's formation process from its beginning, through the cannibalization phase building the large BGG, and finally concluding with a fossil system housing a large BGG and possessing a deficit in bright $L^*$ galaxy members. To better illustrate the hypothesis of formation of a fossil system through the progenitor phase, we have assembled a collage of SDSS images from the CASSOWARY catalog (Figure \ref{fig:collage}).  We order them to simulate the building of a fossil BGG via cannibalization of $L^*$ members.  Early in the progenitor phase, we expect there to be many bright galaxies present in the group and concentrated near the BGG, since dynamical friction has slowed the orbits of the largest galaxies and caused them to fall inward over the group's history.  As time until fossil status is achieved shortens, more and more $L^*$ galaxies will merge with the BGG subsequently shifting the galaxy luminosity function of the group toward a fainter population leaving only one or two bright galaxies near the BGG.  Once the last bright member merges with the BGG, a fossil system will form housing an elongated (possibly asymmetric) BGG.  As the BGG begins to relax after the final major merger it will eventually settle into a massive symmetric elliptical galaxy stereotypical of fossil systems.  A follow up study of progenitors is currently being done using {\it Chandra/HST} data to better see how the hot gas evolves alongside the stellar population as a group draws closer to the fossil threshold (Johnson et al.\, in preparation).  We expect to see a correlation between a progenitor's X-ray properties and time until fossil status is achieved as well.

The high redshift of the CASSOWARY sample is beneficial for locating possible precursors to nearby fossil systems. However, the look back time that allows large BGGs to form also allows new galaxies to fall within $0.5R_{200}$, potentially destroying a system's fossil status before $z=0$.  To account for this chance as conservatively as possible, we define a `danger zone' for each system; this zone is the maximum projected distance from which a galaxy could free-fall inside $0.5R_{200}$ within the look back time.  The free-fall time is given by:
\begin{equation}
t_{ff}=\frac{\pi}{2}\frac{r^{\frac{3}{2}}}{\sqrt{2G(M+m)}}
\end{equation}
where we took $r$ to be the distance from the galaxy down to $0.5R_{200}$ for the group, as this is the threshold for a member to be considered in a system's fossil classification.
In an accelerating universe, the ultimate mass of a galaxy cluster in the far future is around two times the virial mass ($M_{200}$) at $z=0$ (Busha et al.\ 2005).  Since the density inside the virial radius is constant by definition, the virial radius will scale as $M_{200}^{1/3}$ making the final radius $\sim$1.25 times the current $R_{200}$.  The turnaround radius (the point beyond which matter near an overdensity of a certain mass will not collapse inward but expand) is defined as $2R_{200}$, implying that the final turnaround radius will be $\sim2.5R_{200}$ at $z=0$.  However, since our targets are at $z>0$, their masses have grown between when we have observed then and now.  Assuming a mass growth of a factor of four between then and now, that gives a $z=0$ virial radius of $1.6R_{200}^{obs}$ and a $z=0$ turnaround radius of $3.2R_{200}^{obs}$.  For this work, we adopt a turnaround radius of $3.0R_{200}$ and set this as the upper limit to our `danger zone.'
All bright galaxies within the `danger zone' that passed our group inclusion criteria and were bright enough to violate either the Jones ($\Delta m_{12}\geq2.0$) or Dariush ($\Delta m_{14}\geq2.5$) fossil criteria were flagged.  Groups with flagged galaxies within their `danger zone' may still be fossil progenitors, however one cannot rule out the possibility that one or more of the flagged galaxies will eventually fall into the group potential.  We found that out of the 28 strong lensing fossil/progenitor systems in the CASSOWARY catalog, only two (CSWA 26 and CSWA 159) have no nearby galaxies bright enough to endanger their eventual fossil statuses making these true fossil progenitors\footnote[7]{It is interesting to note that both CSWA 26 and CSWA 159 are classified as Dariush fossils and Jones progenitors, implying that in both cases we are witnessing the formation of extremely massive BGGs compared to their respective group richnesses.}.


In simulations, the entire lifetime of fossil systems can be chronicled by observing when and if bright galaxies fall into the group.  Observationally, it is more difficult, as we do not know the proper motions of all galaxies around the group.  Spectroscopic redshifts of galaxies in and near the group can constrain the radial velocities, however the tangential components remain unknown.  This means we cannot know for certain which bright galaxy outside $0.5R_{200}$ will fall inwards.  Therefore, the result that only $\sim$$7\%$ of $0.2<z<0.7$ fossil/progenitor systems will stay fossils until $z=0$ is an extremely conservative estimate.  Spectra of observed fossils and progenitors at higher redshifts have the potential to increase this estimate, as some bright galaxies will undoubtedly be eliminated, being identified as either foreground or background.






\section{Summary}
The progenitors to today's fossil systems have been shown to exist in the universe and can be located.  Kanagusuku et al.\ (2016) found that in the Millennium simulation most of today's fossils finished forming their BGGs between $0.3<z<0.6$ which also happens to be the optimal distance to see strong gravitational lensing.  The discovery of the Cheshire Cat fossil group progenitor (CSWA 2) in the CASSOWARY catalog of strong lensing events in SDSS prompted us to analyze the remaining 57 CASSOWARY members to see if more progenitors could be located.

In this study of CASSOWARY strong lensing systems, we find fossil rates of $13.5\pm2.8\%$ and $17.3\pm2.6\%$ for the Jones et al.\ (2003) and Dariush et al.\ (2010) criteria respectively which is consistent with the expected rate of 8\% -- 20\% of all groups being fossil systems (Jones et al.\ 2003).  This contrasts our non-lensing control fossil rates of $2.9\pm1.6\%$ and $13.6\pm1.2\%$ indicating the presence of a strong lensing bias toward classical (Jones) fossil formation.  Our CASSOWARY progenitor rates of $23.1\pm2.5\%$ and $28.9\pm2.5\%$ (Jones and Dariush respectively) are also elevated compared to our non-lensing progenitor rates of $17.5\pm1.2\%$ and $25.2\pm1.1\%$.  Average galaxy luminosity functions for each class of system (normal, progenitor, and fossil systems) confirmed that fossil progenitors fall between the normal and fossil fits, indicating the formation of the $L^*$ galaxy deficit observed in fossil systems.  For the CASSOWARY sample, the progenitor luminosity function showed a slight deficit of intermediate member galaxies supporting the hypothesis that fossil BGGs are formed via cannibalization of their $L^*$ neighbors and that we are witnessing this process in some CASSOWARY fossil progenitors. 

A control sample of non-lensing groups at similar redshifts and galaxy counts was complied to compare against the CASSOWARY lensing sytems to see how the conditions leading to a strong gravitational arc near the BGG could bias the sample.  It was found that while being a strong gravitational lens did not appear to affect the odds of a group being a fossil progenitor, it does increase the odds that a group will be a Jones fossil, indicating the existence of a strong lensing bias possibly linked to the inital formation of the lensing group.  Comparing the cumulative galaxy luminosity functions of the non-lensing control sample to the CASSOWARY groups showed the non-lensing progenitor function agreeing more with the non-lensing normal groups rather than transitioning to the non-lensing fossil luminosity function.  This could also be an indication of the strong lensing bias preferentially selecting the most extreme examples of fossil formation (i.e., systems with the highest mass concentration, largest intermediate mass galaxy deficit, largest BGGs).  Additionally, we observe lensing systems possessing an average of $2\sigma$ more bright galaxies than the best fit gives for galaxies outside $0.5R_{200}$.  This is not seen in the non-lensing systems, further supporting the existence of a strong lensing bias toward classical (and possibly older) fossil-like systems.

Within our sample of lensing fossil progenitors, we identified fossil progenitors which could better explain the origin of both cool core and non-cool core fossil systems observed today.  Most fossil progenitors in this study seem to be in the process of forming a massive BGG via $L^*$ member cannibalization making these progenitors of the observed cool core fossil population seen today.  We also find two fossil progenitors that show evidence of being a group-group merger (e.g. the Cheshire Cat and CSWA 26).  When the multiple BGGs merge, a fossil system will be created, however the shocked gas will still be cooling a few Gyrs after the galaxy merging is complete (Irwin et al. 2015) making the Cheshire Cat and CSWA 26 progenitors to the observed non-cool core fossil population seen today.  We are engaged in futher work on this topic using {\it Chandra} imaging of the hot gas of eight fossil progenitors in the CASSOWARY catalog at a range of stages in their evolution toward fossil status, from 100 Myr to 5 Gyr until sufficient merging has concluded to establish the required $r$-band magnitude gap, the goal being to observe the evolution of the hot gas component of a fossil progenitor and whether or not a cool core is present as the BGG forms (Johnson et al.\ in preparation).




\acknowledgments
Acknowledgements:  We thank William Keel, Jeremy Bailin, Yuanyuan Su MacLellan, Eddie Johnson, and Spring Johnson for useful conversations and constructive feedback.  This work was supported by {\it Chandra} grant GO6-17106X and {\it HST} grant HST-GO-14362.003-A.

\begin{footnotesize}

\end{footnotesize}

\end{document}